\documentclass[%
superscriptaddress,
reprint,
 amsmath,amssymb,aps, 
 floatfix, showpacs,
prb,
citeautoscript
]{revtex4-1}

\usepackage{graphicx}
\usepackage{dcolumn}
\usepackage{bm}
\usepackage{amsfonts,amssymb,amsmath}
\usepackage{hyperref}
\usepackage{braket}
\usepackage{color}

\newcommand{\ua}{\uparrow}
\newcommand{\da}{\downarrow}

\begin{document}

\preprint{}

\title{Symmetries and hybridization in the indirect interaction between magnetic moments in MoS$_{2}$ nanoflakes}

\author{Oscar \'Avalos-Ovando}
 \email{oa237913@ohio.edu}
 \affiliation{Department of Physics and Astronomy, and Nanoscale and Quantum
 	Phenomena Institute, \\ Ohio University, Athens, Ohio 45701--2979, USA}
\author{Diego Mastrogiuseppe}
\affiliation{Instituto de F\'isica Rosario (CONICET), 2000 Rosario, Argentina}
\author{Sergio E. Ulloa}
\affiliation{Department of Physics and Astronomy, and Nanoscale and Quantum
	Phenomena Institute, \\ Ohio University, Athens, Ohio 45701--2979, USA}

\date{\today}

\begin{abstract}
We study the Ruderman-Kittel-Kasuya-Yosida interaction between magnetic
impurities embedded in $p$-doped transition metal dichalcogenide triangular flakes. The role of underlying symmetries is exposed by analyzing the interaction as a function of impurity separation along zigzag and armchair trajectories, in specific parts of the sample.
The large spin-orbit coupling in these materials produces strongly anisotropic interactions, including a Dzyaloshinskii-Moriya component that can be sizable and tunable.
We consider impurities hybridized to different orbitals of the host transition-metal and identify specific characteristics for onsite and hollow site adsorption. In the onsite case, the different components of the interaction have similar magnitude, while for the hollow  site, the Ising component dominates.
We also study the dependence of the interaction with the level of hole doping, which supplies a further degree of tunability.
Our results could provide ways of controlling helical long range spin order in magnetic impurity arrays embedded in these materials.
\end{abstract}

\pacs{75.30.Hx,75.75.-c,75.70.Tj} 

\maketitle


\section{Introduction}
\label{sec:introduction}

Spintronics relies on the manipulation of the electron spin in materials. Metals or semiconductors with strong
spin-orbit coupling (SOC), such as the layered transition-metal dichalcogenides
(TMDs) \cite{Novoselov2005,Geim2013,Wang2012,Bhimanapati2015}, provide very promising opportunities\cite{Zibouche2014,Han2016}.
When  exfoliated down to a fundamental stack of three atomic layers (which we refer to as \emph{monolayer} from now on), TMDs display rich electronic and optical properties \cite{Xu2014NatPhys, Liu2015, Castellanos2016, Zhu2011,Cheiwchanchamnangij2012,Xiao2012}. MoS$_{2}$, WSe$_{2}$, and WS$_{2}$, are among the most studied TMDs, all exhibiting a direct optical gap in the monolayer limit \cite{Mak2010}. The process of sample production, such as mechanical exfoliation or chemical vapor deposition, often produces nanoscale crystals --nanoflakes-- with different shapes and boundaries, such as stars \cite{Van2013}, hexagons \cite{Cao2015hexagonalflakes},
rhomboids, \cite{Wang2013rhomboidflakes} and
triangles \cite{Lauritsen2007,Van2013,Chiu2014}. The different shapes and boundaries can have a large impact on the properties of
the system. For instance, MoS$_{2}$ zigzag-edge nanoribbons exhibit unusual ferromagnetic properties \cite{Li2008,Botello2009,Tongay2012}, and small-flake polycristalline MoS$_{2}$ films are reported to exhibit intrinsic magnetism \cite{Lauritsen2007}.

A particular form of magnetic interaction takes place when localized magnetic moments in metals interact effectively through an indirect exchange process mediated by the conduction electrons, known as the Ruderman-Kittel-Kasuya-Yosida (RKKY) interaction \cite{RudermanKittel1954,Kasuya1956,Yosida1957}. Even though TMDs are semiconductors, they can be doped with different atomic species to achieve conducting character.
Hole doping is particularly important because the SOC produces a large spin splitting in the valence band near the band edge. Thus the effects of SOC on different physical properties should be more noticeable and controllable in this energy region.
It has been found that \emph{p}-doping of MoS$_{2}$ can be achieved by substituting Mo for Nb \cite{Laskar2014, Suh2014}, with phosphorus implantation \cite{Nipane2016}, and also predicted in ab-initio calculations for different dopants \cite{Dolui2013,Mishra2013,Cheng2013}.
Other materials, such as WSe$_2$, have an intrinsic $p$-type doping.
Localized magnetic moments can be intrinsic to the sample production process or can be introduced extrinsically, for instance, by implantation with an STM tip \cite{Khajetoorians2012, Lounis2014}. This method provides a controlled way of designing magnetic nanostructures. In the case of TMDs, the local moment formation with magnetic dopants has been analyzed by ab-initio studies \cite{Mishra2013,Cong2015,Lu2014,Saab2016}, and in experiments \cite{Zhang2015,Wang2016}.

The RKKY interaction is well understood in conventional metals. However, materials with more complex band structure, with orbital degrees of freedom and strong SOC such as the TMDs, provide a more complex scenario in which the interplay of the various components can give rise to interesting features. In bulk TMD monolayers, a sizable Dzyaloshinskii-Moriya (DM) interaction appears in the indirect exchange, with magnitudes that are comparable to the typical Heisenberg terms \cite{Parhizgar2013,Hatami2014,Mastrogiuseppe2014,Avalos2016,Avalos2016arxiv}.
In general, the details of the hybridization of the magnetic species with the local host, as well as the size of the system, have large impact on the effective interaction between impurities, such as in two-dimensional (2D) electron gas nanoribbons \cite{Mi2011}.

In a 2D lattice, the magnetic moments can  hybridize in different ways. The most common places are: on top of a lattice site (onsite), on the line between two lattice sites (bridge), in hollow sites (plaquette), or substitutional. The onsite hybridization has been studied extensively in infinite graphene \cite{Power2013,Kogan2011},  nanoflakes \cite{Szalowski2011,Szalowski2013,Nikoofard2016}, nanoribbons \cite{Szalowski2013jofcm,Black2010paper1,Akbari2014}, and also in infinite TMD layers \cite{Parhizgar2013,Hatami2014,Mastrogiuseppe2014} and flakes \cite{Avalos2016,Avalos2016arxiv}. The plaquette configuration has been analyzed in 2D graphene \cite{Saremi2007,Black2010paper1,Uchoa2011,Sherafati2011},  triangular flakes \cite{Szalowski2013,Nikoofard2016}, and carbon nanotubes \cite{Kirwan2008,Gorman2015}. The effective interaction has been also studied in other systems with large intrinsic SOC, such as silicine \cite{Xiao2014,Zare2016}, and Pt lattices \cite{Patrone2012}.

Finite TMD samples exhibit highly localized states near the edges of the flake \cite{Bollinger2001,Pavlovic2015,Segarra2016,Farmanbar2016,Rostami2016}, resulting in noncolinear and tunable long range interactions when the impurities sit at these edges, and with slow decay with the impurity separation
\cite{Avalos2016,Avalos2016arxiv}. The plaquette hybridization geometry has not yet been reported on TMDs.

In this paper, using an effective three-orbital tight-binding model \cite{Liu2013} that captures the relevant bands and symmetries at low energies, we study the interaction between two magnetic impurities in \emph{p}-doped triangular TMD nanoflakes, for both onsite and plaquette configurations. In the onsite configuration, the impurities hybridize on top of single transition-metal atoms, while in the plaquette case they sit in hollow sites of transition-metal triangles, as we will describe in detail. We analyze the effective exchange interaction as a function of the impurity separation, comparing the behavior of impurities on the edges to the ones in the bulk of the flake. We find that both the onsite and plaquette configurations display helical couplings, with sizable Dzyaloshinskii-Moriya interaction.
Interestingly, the plaquette configuration shows a larger Ising interaction compared to the in-plane terms, which is explained by second order perturbation theory calculations.
We also find that the interaction depends strongly on the direction of impurity separation, either zigzag or armchair, highlighting the importance of crystal symmetries in the effective exchange. We further analyze the possible tunability of the strength and anisotropy of the interaction with the doping concentration, and identify different scattering processes that contribute to the effective coupling.

\section{Model and Approach}
\label{sec:model}

We focus on triangular zigzag-terminated MoS$_{2}$ nanoflakes\cite{Van2013}, with two magnetic moments (or impurities) hybridized to different lattice environments, including onsite and plaquette  (or hollow) configurations.
\begin{figure*}
  \centering
  \includegraphics[width=0.5\textwidth]{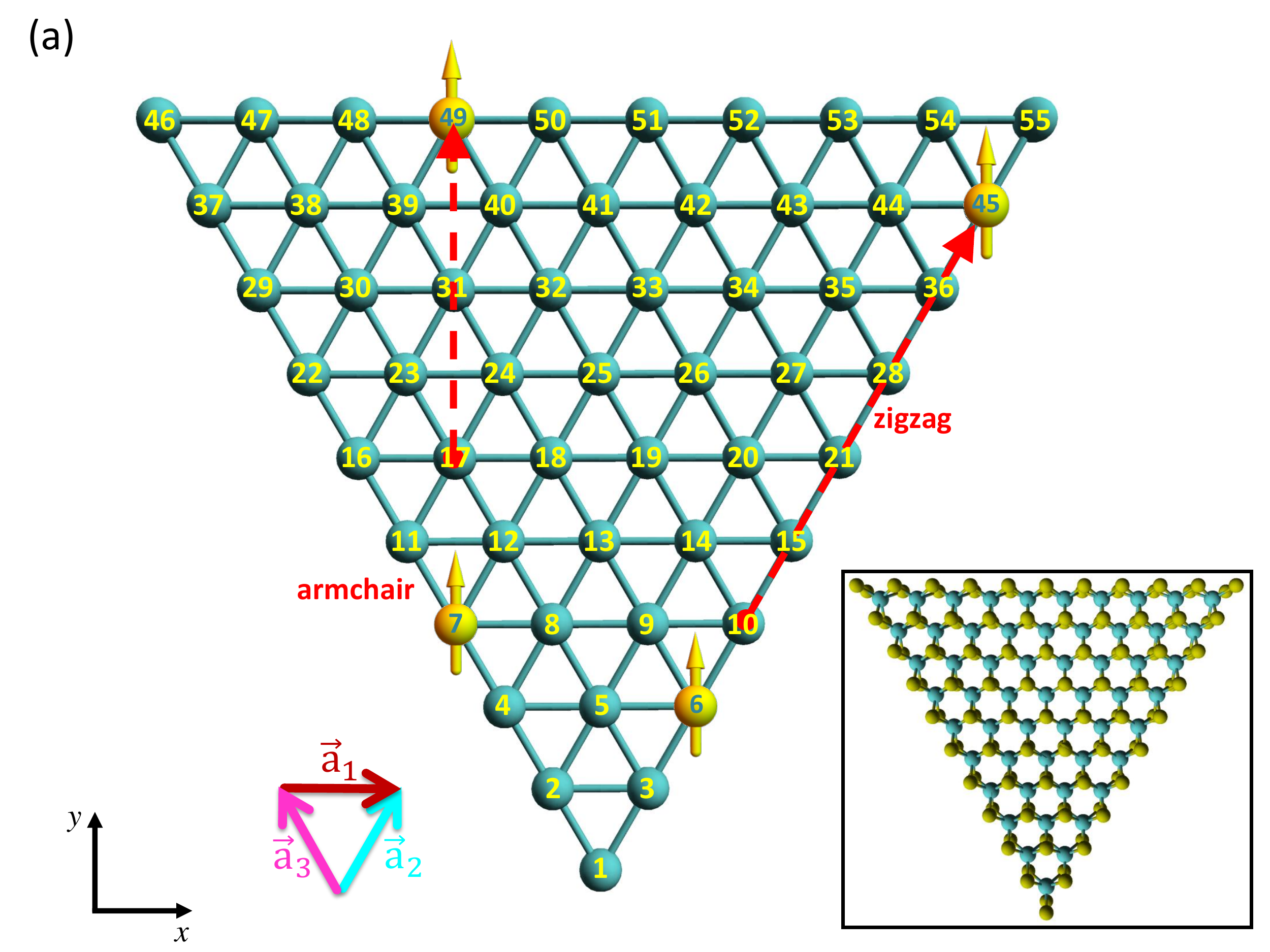}\includegraphics[width=0.25\textwidth]{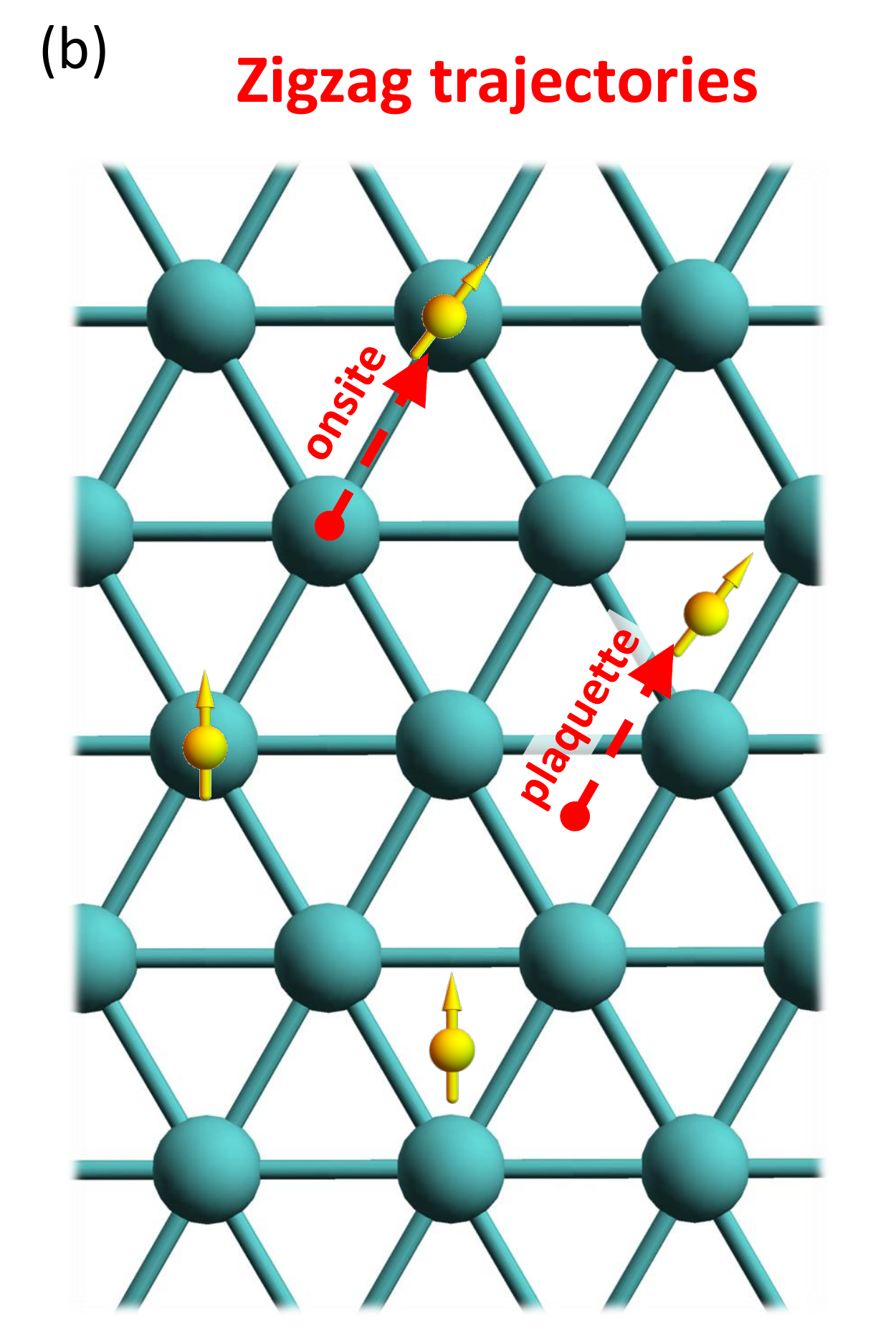}\includegraphics[width=0.25\textwidth]{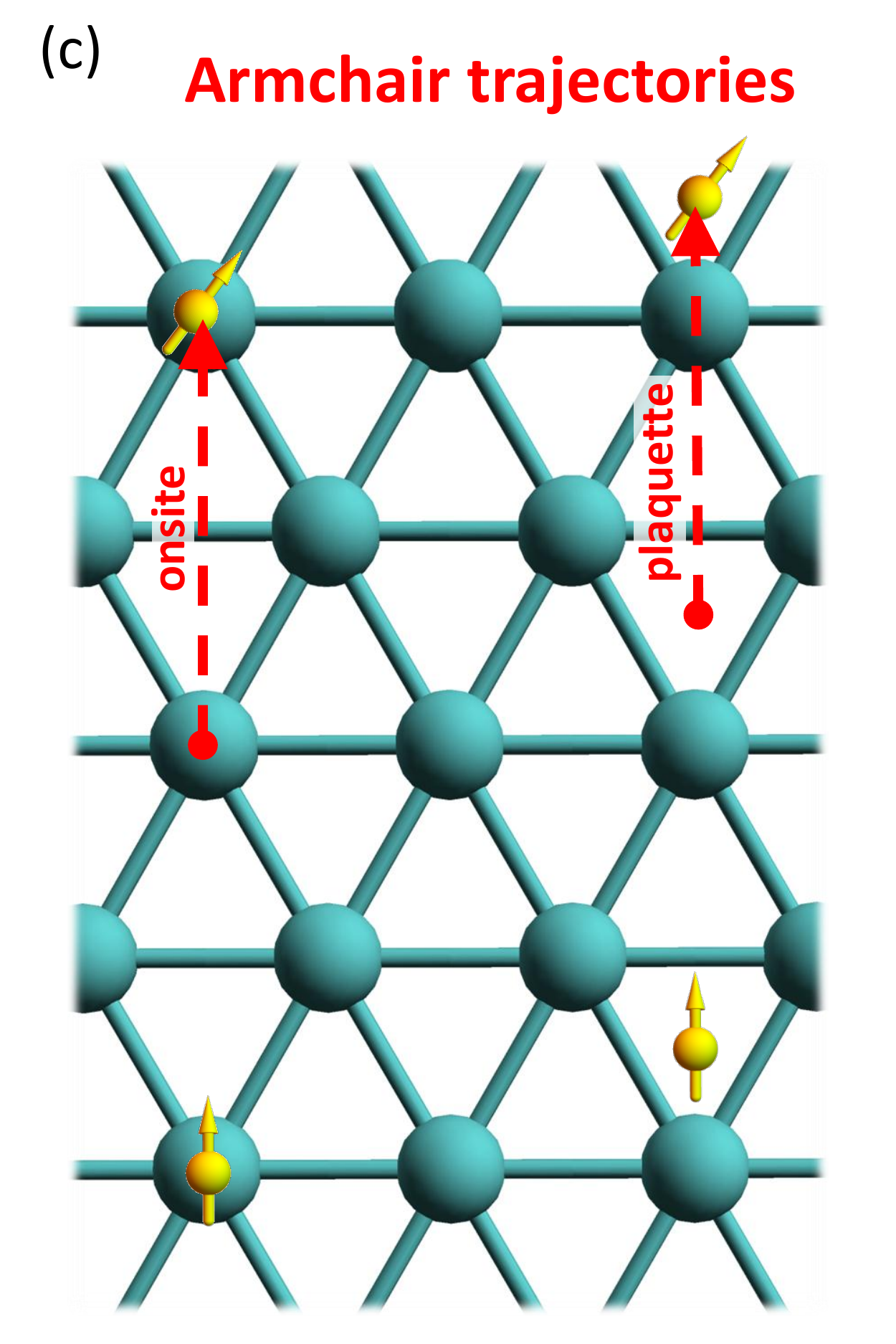}\\
  \caption{(Color online) (a) Top view of the effective lattice used to simulate the triangular zigzag-terminated TMD nanoflake. Each site represents a Mo atom, including the numbering used to construct the flake. The inset shows a top view of the real representation of the flake, with the Mo (S) atoms shown in dark green (dark yellow). The three hopping directions are given by $\bm{a}_{1}=a(1,0)$, $\bm{a}_{2}=a(1/2,\sqrt{3}/2)$ and $\bm{a}_{3}=\bm{a}_{2}-\bm{a}_{1}$, where $a$ is the lattice constant. Yellow arrows represent one pair of magnetic moments in the zigzag direction and another pair in the armchair direction. One impurity is held fixed and the other is moved along the corresponding direction, as indicated by red dashed lines.  Two independent zigzag (b) and armchair (c) trajectories for both onsite and plaquette \emph{triangle down} configurations.}\label{fig1}
\end{figure*}
The host material can be described by a triangular lattice of Mo atoms since, at low energies, only three 4\emph{d}-orbitals from these atoms contribute significantly \cite{Liu2013} (see Fig.\ \ref{fig1}). We use a three-orbital tight-binding model, with $d_{z^2}$, $d_{xy}$, and $d_{x^2-y^2}$ Mo orbitals. The full  Hamiltonian is given by
\begin{equation}\label{hamiltonianwithimpurities}
  H = H_{0} + H_{\text{I}},
\end{equation}
where $H_{0} = H_{\text{on}} + H_{\text{t}}$ (onsite + hoppings) describes the TMD without impurities, and $H_{\text{I}}$ models the interaction of two magnetic impurities  with the conduction electrons of the host.
The onsite Hamiltonian is given by
\begin{equation}\label{eq:onsite}
  H_{\text{on}} = \sum_{j=1}^{N_t} \sum_{s=\ua,\da} \sum_{\alpha,\alpha'} \varepsilon_{\alpha,\alpha',s}\:d_{\alpha,s}^{\dagger}(\bm{r}_j) d_{\alpha',s}(\bm{r}_j),
\end{equation}
where $d_{\alpha,s}(\bm{r}_j)$ [$d^{\dagger}_{\alpha,s}(\bm{r}_j)$]
annihilates [creates] a spin-$s$ electron at the lattice site
$\bm{r}_j=j_1 \bm{a}_1 + j_2\bm{a}_2$ and orbital $d_\alpha$. The $\bm a_l$ are lattice vectors with lattice constant $a$ (Fig.\
\ref{fig1}), $\alpha \in\,\left\{z^2\equiv 0, xy\equiv 1, x^2-y^2\equiv 2\right\}$, and
 $\varepsilon_{\alpha,\alpha',s}$ are the onsite energies.
The total number of sites in the sample, $N_t$, is given by the number of rows or atoms on the edge $N_e$, as $N_t=N_e(N_e+1)/2$.
The  hopping Hamiltonian $H_{\text{t}}$ is given by
\begin{equation}\label{eq:hopping}
H_{\text{t}} = \sum_{j,s,\alpha,\alpha'} \sum_{l=1}^{3} t_{\alpha,\alpha'}^{(\bm{a}_{l})}\,d_{\alpha,s}^{\dagger}(\bm{r}_j) d_{\alpha',s}(\bm{r}_j+\bm{a}_l) + \text{H.c.},
\end{equation}
where the $t_{\alpha,\alpha'}^{(\bm{a}_{l})}$ are the orbital-dependent hopping parameters in the three nearest-neighbor directions $l=1,2,3$. The different onsite energies and hopping parameters are taken from Refs.\ \onlinecite{Liu2013, Pavlovic2015}, and reproduced in Table\ \ref{tab:table1}.
\begin{table*}[hbt]
	\caption{\label{tab:table1}Onsite $\varepsilon_{\alpha,\alpha',s}$ and hopping $t_{\alpha,\alpha'}^{(\bm{a}_{l})}$ tight-binding  energy parameters for MoS$_{2}$ (taken from Refs.\ \onlinecite{Liu2013, Pavlovic2015}), for directions $\bm a_l$  and  orbitals pairs $d_\alpha,d_{\alpha'}$, with $\alpha, \alpha' \in\,\left\{z^2, xy, x^2-y^2\right\}$. All the energies in eV.}
	\begin{ruledtabular}
		\begin{tabular}{cccccccccc}
			&\multicolumn{9}{c}{$\alpha,\alpha'$}\\ \cline{2-10}\\
			Parameter&$z^2,z^2$ & $z^2,xy$&$z^2,x^2-y^2$&$xy,z^2$&$xy,xy$&$xy,x^2-y^2$&$x^2-y^2,z^2$&$x^2-y^2,xy$&$x^2-y^2,x^2-y^2$\\ \hline
			&&&&&&&&&\\
			$\varepsilon_{\alpha,\alpha',\uparrow}$&1.046&0&0&0&2.104&0.073$i$&0&-0.073$i$&2.104\\
			&&&&&&&&&\\
			$\varepsilon_{\alpha,\alpha',\downarrow}$&1.046&0&0&0&2.104&-0.073$i$&0&0.073$i$&2.104\\
			&&&&&&&&&\\
			$t_{\alpha,\alpha'}^{(\bm{a}_{1})}$&-0.184&0.401&0.507&-0.401&0.218&0.338&0.507&-0.338&0.057\\
			&&&&&&&&&\\
			$t_{\alpha,\alpha'}^{(\bm{a}_{2})}$&-0.184&0.640&0.094&0.239&0.097&-0.268&-0.601&0.408&0.178\\
			&&&&&&&&&\\
			$t_{\alpha,\alpha'}^{(\bm{a}_{3})}$&-0.184&-0.640&0.094&-0.239&0.097&0.268&-0.601&-0.408&0.178\\
		\end{tabular}
	\end{ruledtabular}
\end{table*}
$H_0$ can be diagonalized by a change of basis
\begin{equation} \label{eq:diagonaliz}
d_{\alpha,s}(\bm{r}_j) = \sum_{\mu=1}^{3 N_t} \psi_{k,\mu,s}\, c_{\mu,s},
\end{equation}
such that
\begin{equation}
H_0 = \sum_{\mu=1}^{3N_t}\sum_s \varepsilon_\mu c^\dagger_{\mu,s} c_{\mu,s},
\end{equation}
where $k=3j-2+\alpha$, such that $\psi_{k,\mu,s}$ is the $\mu$th component of the eigenvector for site $j$, orbital $\alpha$, and spin projection $s$. As the TMD Hamiltonian does not mix spin, each spin block can be diagonalized separately. Due to time reversal symmetry, we have that $\psi_{k,\mu,\ua}\equiv \psi_{k,\mu}=\psi_{k,\mu,\da}^*$.
Here, we have assumed that the original (spin up block) basis is arranged as $[d_{0,\ua}(\bm{r}_1), d_{1,\ua}(\bm{r}_1), d_{2,\ua}(\bm{r}_1), \cdots , d_{0,\ua}(\bm{r}_{N_t}), d_{1,\ua}(\bm{r}_{N_t}),\\ d_{2,\ua}(\bm{r}_{N_t})]^T$ and the diagonal one as $[c_{1,\ua}, c_{2,\ua},\cdots,c_{3N_t,\ua}]^T$, in ascending order of eigenvalues $\varepsilon_\mu$. In order to simplify the notation, we define $\psi^{z^2}_{j,\mu} \equiv \psi_{3j-2,\mu}$, $\psi^{x^2-y^2}_{j,\mu} \equiv \psi_{3j-1,\mu}$, and $\psi^{xy}_{j,\mu} \equiv \psi_{3j,\mu}$.

\begin{figure}[htb]
  \centering
  \includegraphics[width=0.47\textwidth]{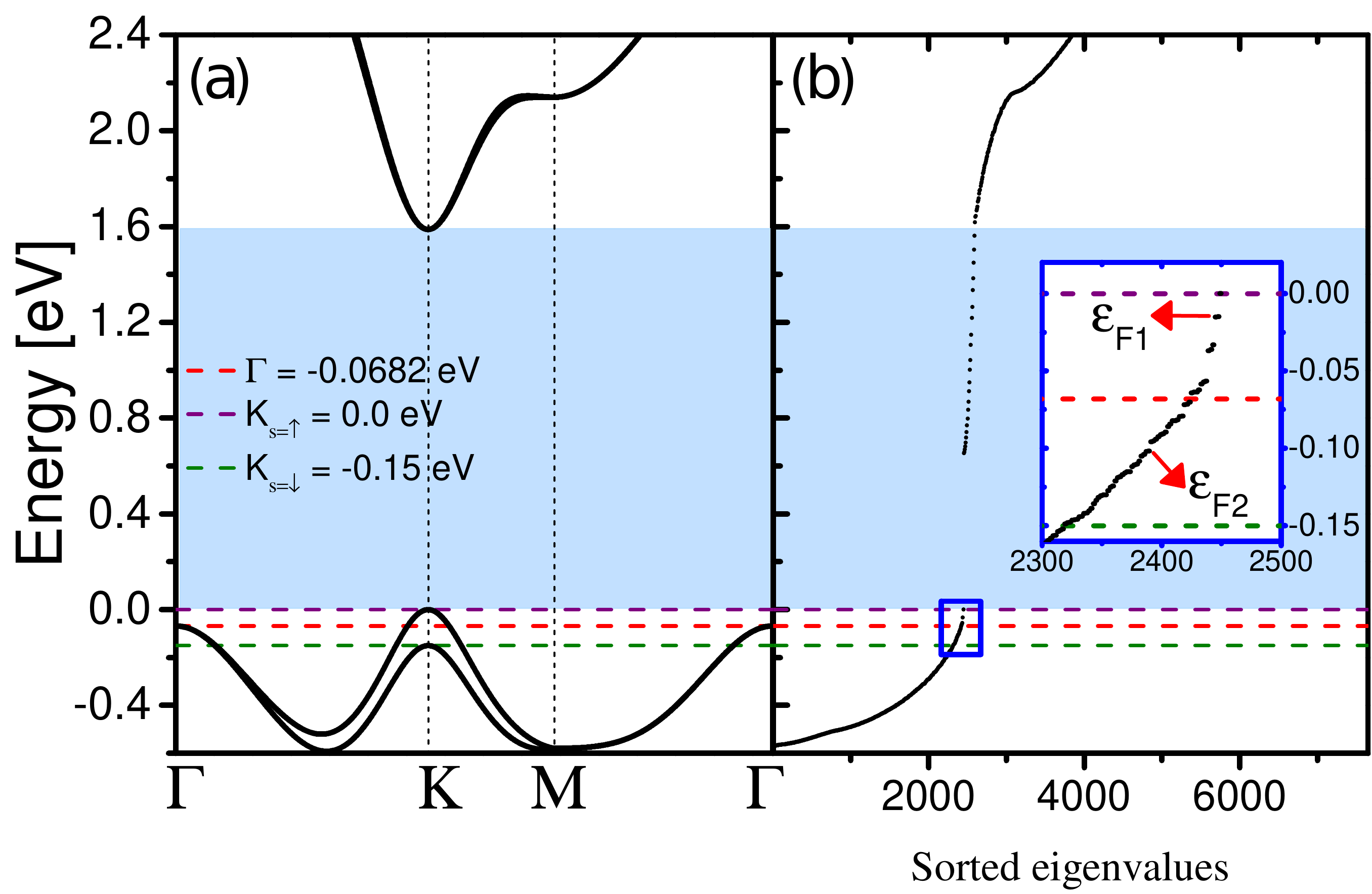}
  \caption{(Color online) (a) High symmetry directions in the first Brillouin
  zone of the infinite MoS$_{2}$ monolayer. The valence band maximum (VBM) at
  $K$ is shifted to zero energy and the energy levels of the $\Gamma$ point,
  and $K$ for spin up and down are shown in dashed lines. The light blue area
  indicates the direct gap ($\sim1.6$ eV). (b) Discrete energy levels for a
  50-row flake. Edge states generated by the finite size appear in the gap.
  Inset shows states near the VBM. $\varepsilon_{F1}$ and $\varepsilon_{F2}$ represent the two different levels of doping (or gating) considered in this work.}\label{fig2}
\end{figure}

In the infinite MoS$_{2}$ monolayer, the first Brillouin zone has two inequivalent $K$ and $K'$ points, with a sizable spin splitting around the valence band maximum (VBM), as shown in Fig.\ \ref{fig2}(a). There is a direct band gap ($\sim1.6$ eV) between the VBM and the conduction band minimum (CBM) at these two points, with definite spin-valley relation, due to the absence of inversion symmetry. On the other hand, for finite systems, such as the  triangular flakes studied here, the electronic spectrum is fully discrete, showing both bulk- and edge-like states, as shown in Fig.\ \ref{fig2}(b). States from both the valence and conduction bands have been brought into the gap, corresponding to one-dimensional-like (1D) extended states localized near the borders of the sample.\cite{Bollinger2001,Segarra2016}

Figures \ref{fig1}(b) and (c) show onsite and plaquette connections along  zigzag and armchair trajectories respectively. The Hamiltonian for the magnetic impurities connected to specific sites of the TMD lattice is given by
\begin{equation}\label{impurities1}
  H_{\text{I}}=\sum_{i=1,2} \mathcal{J}_{\alpha_i}\: \bm{S}_{i}\cdot\bm{s}_{\alpha_i}(\bm{r}_i),
\end{equation}
where $\mathcal{J}_{\alpha_i}$ is the exchange coupling between the localized magnetic moment $i$, represented by $\bm{S}_{i}$,  and electron spin density at lattice site $\bm{r}_{i}$ and orbital $\alpha_i$, given by
\begin{equation}\label{impurities2}
  \bm{s}_{\alpha}(\bm{r})=\frac{1}{2}\sum_{s,s'}d_{\alpha,s}^{\dagger}(\bm{r}) \bm{\sigma}_{s,s'}d_{\alpha,s'}(\bm{r}),
\end{equation}
where $\bm{\sigma}$ is the vector of spin-$\frac12$ Pauli matrices. If the impurity is in a plaquette environment, the previous description holds but now, in Eq.\ \ref{impurities1}, one has to sum over the three Mo sites surrounding the impurity as well.

In the bulk 2D crystal, the electronic degrees of freedom can be integrated out using second order perturbation theory and the effective interaction can be obtained analytically  \cite{Mastrogiuseppe2014}. This procedure yields the effective exchange Hamiltonian
\begin{eqnarray}\label{jeffective1}
H_{RKKY} &=& J_{XX}\left(S_{1}^{x}S_{2}^{x}+S_{1}^{y}S_{2}^{y}\right)+J_{ZZ}S_{1}^{z}S_{2}^{z}\nonumber\\
 & &+J_{XY}\left(\bm{S}_{1}\times \bm{S}_{2}\right)_{z},
\end{eqnarray}
where all the effective $J$'s are proportional to the static spin susceptibility tensor of the electron gas\cite{RudermanKittel1954,Kasuya1956,Yosida1957}. The net effective interaction is a competition between Ising $J_{ZZ}$, in-plane parallel $J_{XX}$ (=$J_{YY}$),  and cross $J_{XY}$ Dzyaloshinskii-Moriya (DM) terms. In the TMDs, these spin anisotropies are generated by the strong SOC and the absence of inversion symmetry.

In order to calculate the effective $J$'s in our finite sample, we consider the difference between ground state energies of the electron gas with triplet and singlet configurations of the impurities (hybridized to orbitals $\alpha_1$ and $\alpha_2$ respectively), as \cite{Deaven1991,Black2010paper1}
\begin{equation}\label{jeffective2}
  J_{\beta\beta'}^{\alpha_1,\alpha_2}= 2 \left[E(\ua_{\beta},\ua_{\beta'})-E(\ua_{\beta},\da_{\beta'})\right],
\end{equation}
where $\beta$ ($\beta'$) $\in\{X, Y, Z\}$ represents the direction of the spin projection for the first (second) magnetic impurity. \footnote{Notice that we use capital letters for the spin direction in order to avoid confusion with the notation for orbitals.} For instance, $J_{XY}^{z^2,xy}$ is the interaction strength between impurities when the spin of the first one is pointing in the $X$ direction and is hybridized to a Mo $d_{z^2}$ orbital, whereas the spin of the second one is pointing along $Y$ and is hybridized to a $d_{xy}$ orbital. This non-perturbative approach is valid even for large values of local ${\cal J}$ and is capable of generating results for any hybridization geometry and separation between impurities\cite{Black2010paper1}. Notice that positive [negative] values of $J$ correspond to antiferromagnetic (AFM) [ferromagnetic (FM)] alignment between impurities. The ground state energy of the system, including both impurities in a given spin configuration, is defined as the sum of the sorted energy states of the full Hamiltonian up to the Fermi energy $\varepsilon_{\text{F}}$, as
\begin{equation}\label{groundstate}
  E(\bm{S}_{1},\bm{S}_{2})=\sum_{s,\nu=1}^{\varepsilon_{\text{F}}} E_{\nu,s}.
\end{equation}
These eigenenergies are obtained by exact numerical diagonalization of the full Hamiltonian $H$, described by a matrix  of size $6 N_t\times 6 N_t$. The eigenvalues are sorted in ascending order, such that $E_{\nu,s}\leq \varepsilon_{F}$, to carry out the summation.

\section{Results}
\label{sec:results}

Our triangular MoS$_{2}$ flakes consist of $N_e=50$ rows, corresponding to a total of $N_t=1275$ sites ($\simeq$ 160 $\text{\AA}$ on edge). Midgap states appear because of the finite size, having a majority $d_{z^2}$ character and amplitudes that are strongly localized near the borders of the crystallite. These edge states have clear 1D character with momentum along the edge of the flake\cite{Pavlovic2015,Segarra2016}, and their role mediating the effective exchange interaction between magnetic impurities has been recently explored \cite{Avalos2016,Avalos2016arxiv}. In this work, however, we  focus on the bulk-like states at lower energies, close to the VBM, for two different doping levels represented by Fermi energies $\varepsilon_{F1}=-0.0332$ eV and $\varepsilon_{F2}=-0.1018$ eV, as shown in the inset of Fig.\ \ref{fig2}(b). These doping levels correspond to 106 and 160 holes in the flake, or $9.6\times10^{13}$ and $1.4\times10^{14}$ holes/cm$^{2}$, respectively. Notice that one could also consider an intrinsically \emph{n}-doped flake. However, the splitting of states by the SOC is much smaller ($\approx 3$ meV).

\begin{figure*}[htb]
	\centering
	\includegraphics[width=0.33\textwidth]{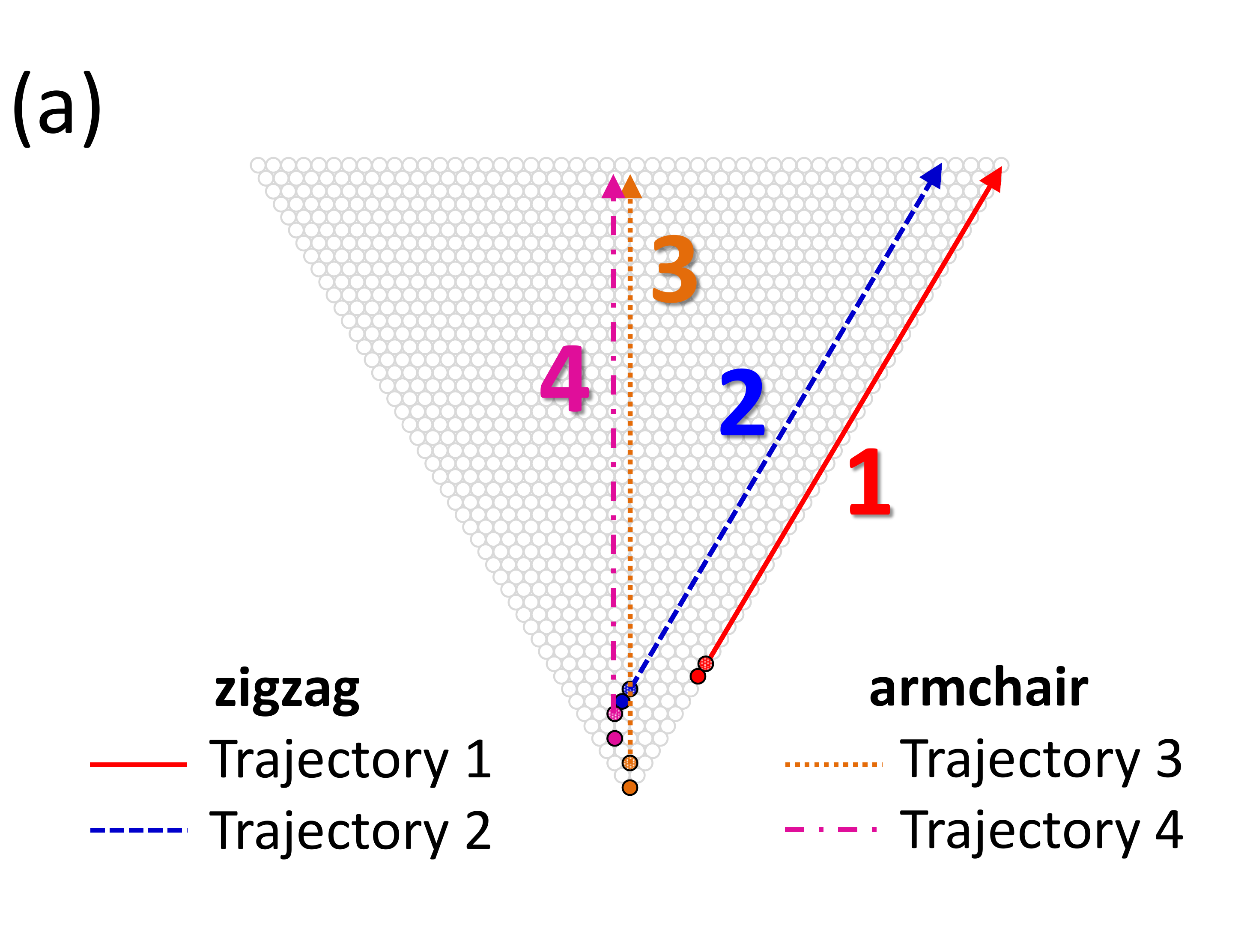}\includegraphics[width=0.33\textwidth]{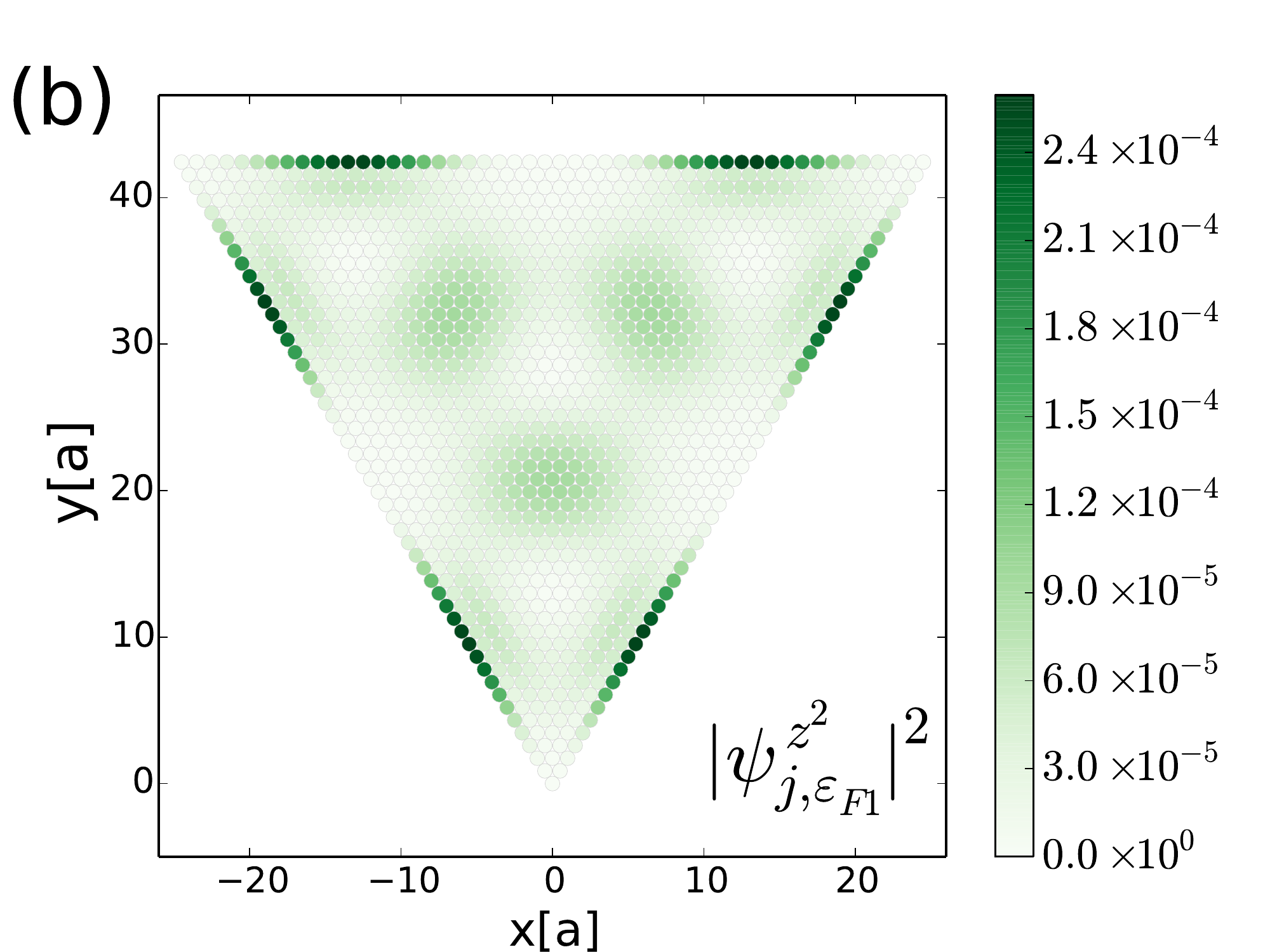}\includegraphics[width=0.33\textwidth]{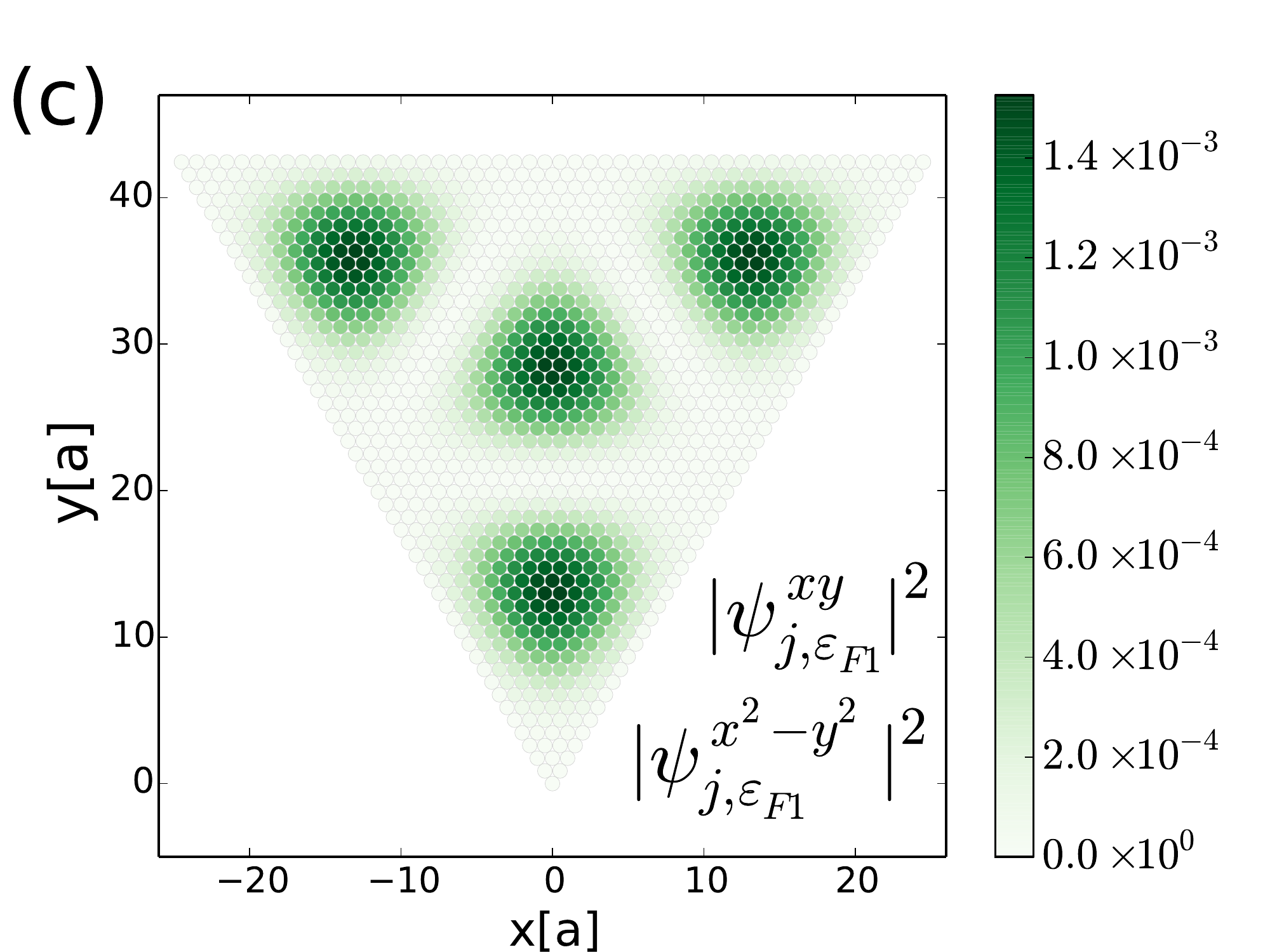}\\
	\includegraphics[width=0.33\textwidth]{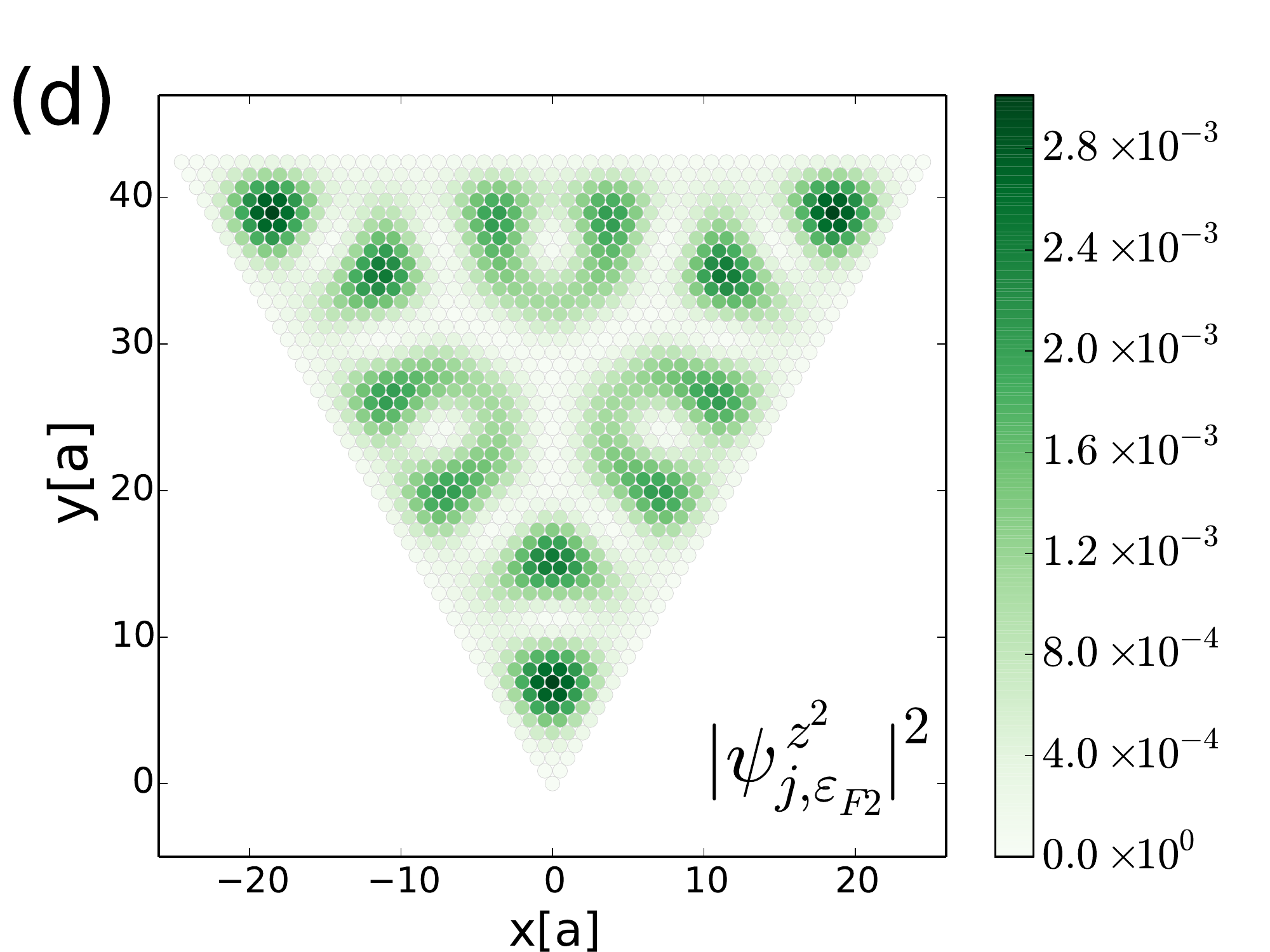}\includegraphics[width=0.33\textwidth]{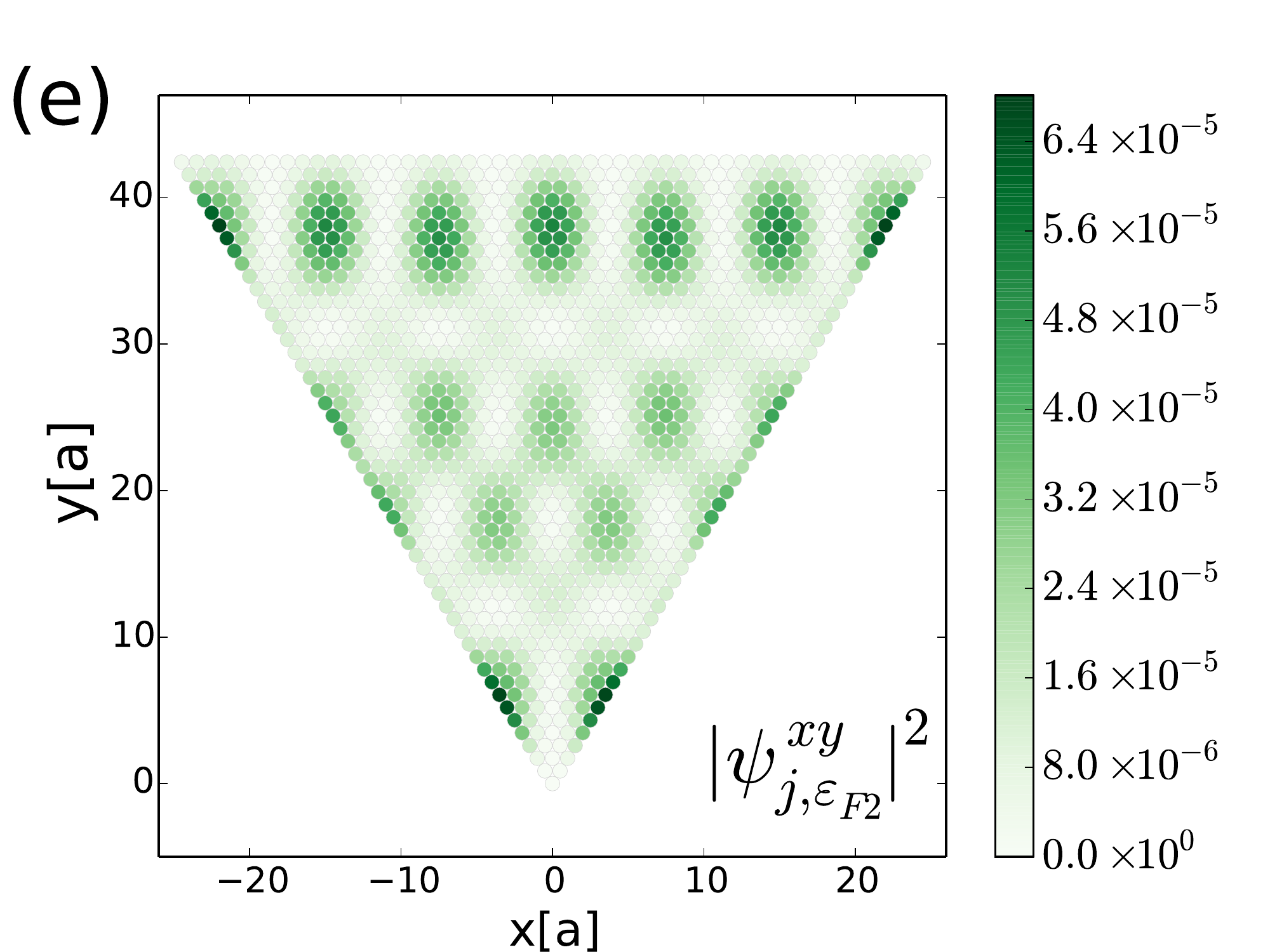}\includegraphics[width=0.33\textwidth]{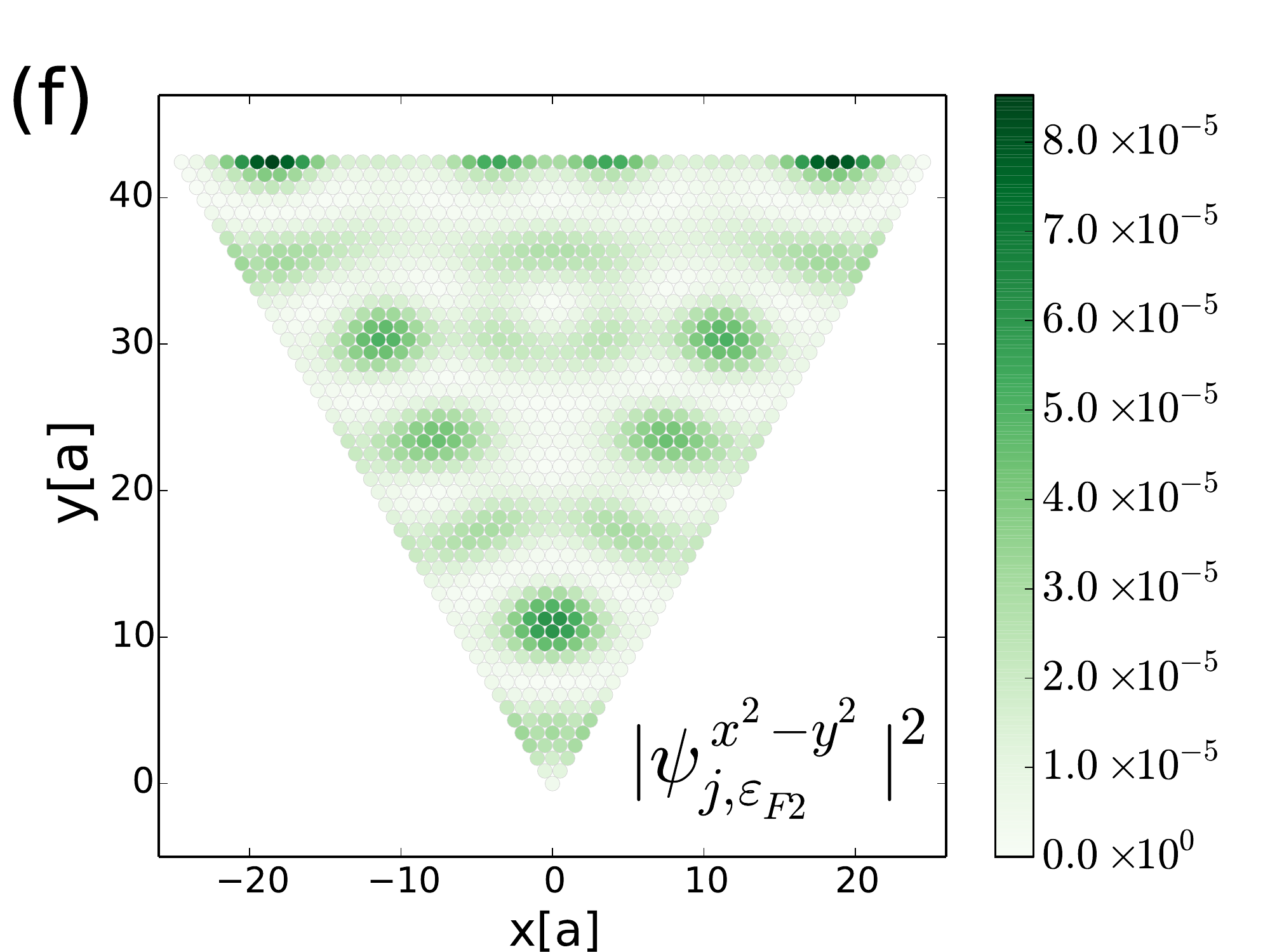}\\
	\caption{(Color online) (a) Four different impurity trajectories for onsite hybridizaton, fixing the first impurity on top of a given atom and moving the second one away from the first. In trajectory 1 (red solid line), the first impurity is located at the 10th row on the edge and the second one is moved along the zigzag direction $\bm{a}_{2}$, starting on the 11th row. Trajectory 2 (dashed blue line) represents a different zigzag direction in the bulk of the flake. For trajectory 3 (orange dotted line), the first impurity is at the bottom corner of the flake, while the second moves up along the armchair direction $\bm{a}_{2}+\bm{a}_{3}$. Trajectory 4 (pink dot-dashed line) is shifted laterally with respect to the previous one. (b)-(f) Orbital-resolved  magnitude squared of the wave function, for two doping levels and the three different orbitals, as indicated in each panel.}\label{fig3}
\end{figure*}

Next, we consider the role of different hybridization environments on the effective exchange interaction between impurities. We focus first on onsite hybridizations in subsection \ref{subsec:onsite}, followed by plaquette environments in subsection \ref{subsec:plaquette}. In each environment, we contrast the behavior at different doping levels, as they contain different orbital and spatial symmetries. In all cases, the first impurity is fixed at a given initial position and the second one is moved along high symmetry directions, as shown schematically in Fig.\ \ref{fig1}(b) and (c). In order to explore boundary effects from   the finite system, we consider two zigzag and two armchair trajectories, as shown in  Fig.\ \ref{fig3}(a). For simplicity, we also consider that the local exchange coupling $\mathcal{J}$ is the same for both impurities, irrespective of the orbital to which they hybridize.

\subsection{Onsite Hybridization}
\label{subsec:onsite}

\subsubsection{Doping level $\varepsilon_{F1}$}

We first set our attention on doping level  $\varepsilon_{F1}$ near the top of the VBM, as seen in Fig.\ \ref{fig2}(b). At this doping level there are no states  from the $\Gamma$ point in the infinite monolayer, thus we expect the states in the flake to have a majority $d_{xy}$ and $d_{x^2-y^2}$ character. In Fig.\ \ref{fig3}(b) and \ref{fig3}(c) we show the normalized wave functions in real space, $|\psi^\alpha_{j,\varepsilon_{F1}}|^2$, for the corresponding unperturbed state. We can see that for orbital $d_{z^2}$ the wave function is mostly localized at the flake edges (as seen in the case of midgap doping levels \cite{Avalos2016,Pavlovic2015}), while for $d_{xy}$ and $d_{x^2-y^2}$ the wave function is symmetric in the $xy$ plane and mostly located inside the flake with much larger amplitudes. Notice that each state is doubly degenerate due to conservation of the spin projection in the pristine flake. The wave functions for each spin are complex conjugates, so the spatial distribution of the  magnitude squared is identical.

\begin{figure}
	\centering
	\includegraphics[width=0.48\textwidth]{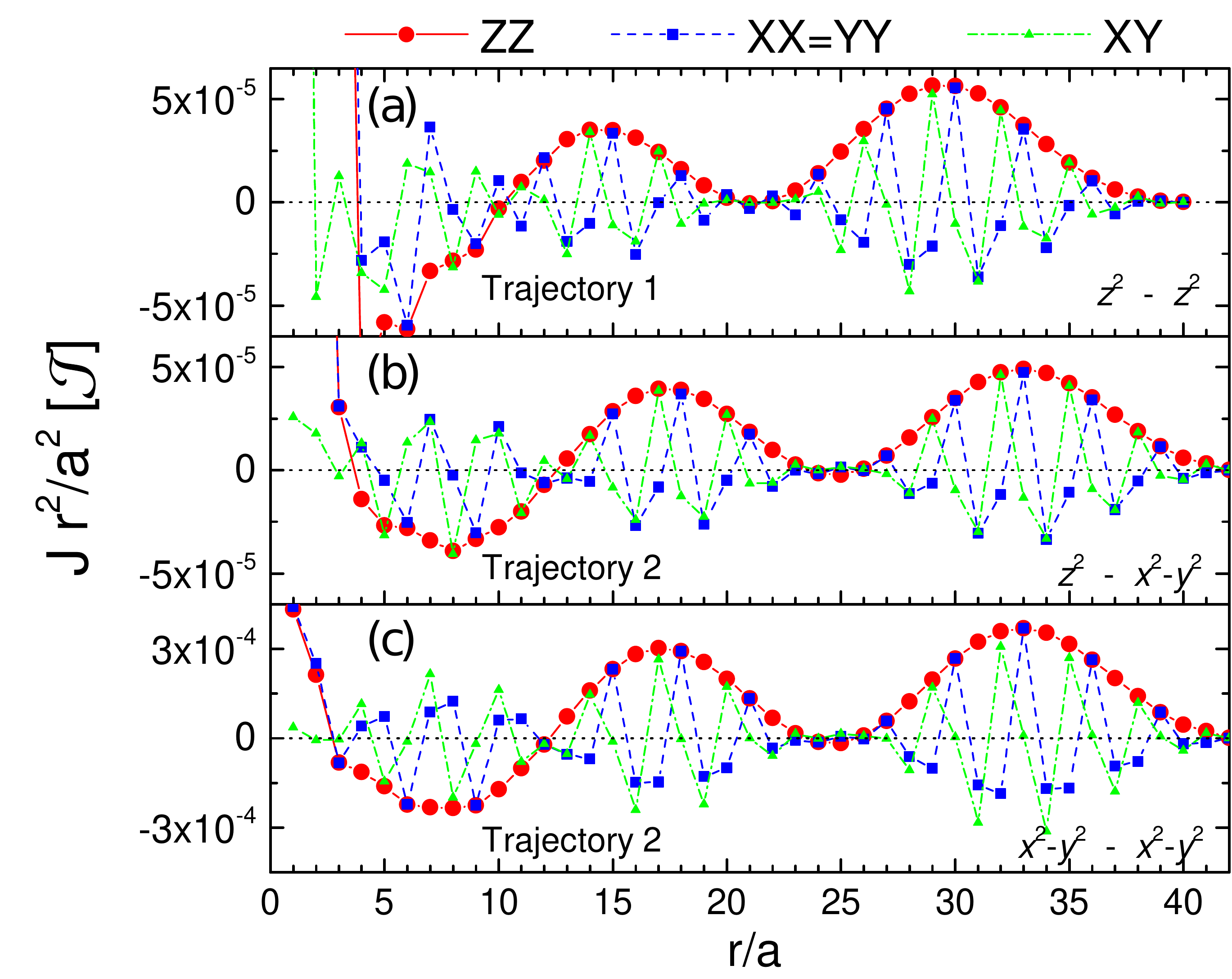}
	\caption{(Color online) The three components of the effective impurity interaction, scaled by $(r/a)^{2}$, versus relative distance along  zigzag directions. All curves correspond to  $\varepsilon_{F1}$, and onsite hybridization to the orbitals indicated in the panels. The different trajectories are explained in Fig.\ \ref{fig3}(a).}\label{fig4}
\end{figure}

Now we analyze the RKKY interaction along \emph{trajectory} 1 on the edge of the flake, as indicated in Fig.\ \ref{fig3}(a), for $\varepsilon_{F1}$. In Fig.\ \ref{fig4}(a) we show the  interaction, in units of ${\cal J}$ and scaled by $(r/a)^{2}$, versus the distance between impurities $r=|\bm r_1 - \bm r_2|$, when both of them are hybridized to $d_{z^2}$ orbitals.
The nearly constant amplitude of the curves indicates a $r^{-2}$ decay, as expected for 2D systems.
 Notice that the Ising component, $J_{ZZ}$,  has a long period of oscillation, of about 15 sites, and $J_{ZZ}>0$ for $r>10$, so that the impurities align mostly AFM for large separations. On the other hand, the parallel and crossed in-plane  interactions $J_{XX}$ and $J_{XY}$ possess a much shorter period of oscillation, about 3 sites, alternating between AFM and FM  as the impurities separate. Also notice that these in-plane interactions have a relative phase difference of nearly one site between them. At specific separations, however, both in-plane interactions are FM (e.g., $r/a=13,28$), while in general they compete against each other.
The interaction along \emph{trajectory} 1 is strong only when one of the impurities is hybridized to a $d_{z^2}$ orbital. We find that $J^{z^2,xy}$ and $J^{z^2,x^2-y^2}$ are typically 10 times smaller than $J^{z^2,z^2}$, but with similar periods of oscillation. On the other hand, hybridizations with in-plane orbitals ($d_{xy}$ with $d_{x^2-y^2}$ and vice versa), produce interactions that are 100 times smaller than $J^{z^2,z^2}$ since, on the edges, these wave functions are nearly negligible (not shown).

In general, we find that the strength of the indirect interaction can be tailored by setting the impurities at points where the modulus squared of the wave function has large amplitudes. However, this should be taken only as a qualitative reference because, in fact, the RKKY interaction is composed of a combination of particle-hole excitations in the electron gas, and it is not directly related to the wave functions of the states at the Fermi level only.


When the impurities are located away from the edges, we notice qualitative changes. Along \emph{trajectory} 2, the wave functions $d_{xy}$ and $d_{x^2-y^2}$ are large in magnitude, but $d_{z^2}$ is negligible. The interaction shows the same modulation as that on the edge, i.e. a large period for $J_{ZZ}$ and a short one for the in-plane terms, but with amplitudes that depend on orbital hybridization. For $J^{z^2,x^2-y^2}$ [Fig.\ \ref{fig4}(b)], or $J^{z^2,xy}$, the interaction is of the same order as that on the edge. When both impurities are hybridized to  $d_{x^2-y^2}$ [Fig.\ \ref{fig4}(c)], or $d_{xy}$, the largest interaction is nearly 10 times larger than that on the edge. When the first impurity is connected to $d_{x^2-y^2}$ or $d_{xy}$, and the second to $d_{z^2}$, the in-plane interactions oscillate as expected, but the slow varying envelope provided by $J_{ZZ}$ shows here a rather weak modulation (not shown), associated with the rather constant (and small) amplitude of $d_{z^2}$ in this internal region of the flake.

We can see from Figs.\ \ref{fig4}(a-c), that the Ising $J_{ZZ}$ effective interaction shows a longer oscillation period than the parallel $J_{XX}$ and DM $J_{XY}$ in-plane interaction terms. This behavior can be explained from the different intra-(for $J_{ZZ}$) and inter-valley (for $J_{XX}$ and $J_{XY}$) scattering processes dominating the interaction. $J_{ZZ}$ is dominated by processes that occur within the same $K$ or $K'$ valley, where no spin flips are allowed in the scattering processes. In $J_{XX}$ and $J_{XY}$, the short period is due to intervalley processes that occur when the electron scatters from $K$ to $K'$ or $\Gamma$ (and vice versa), together with a spin flip. Interestingly, we observe a beating pattern in the in-plane terms with the Ising term acting as the envelope. The details of the oscillation periods naturally depend on the Fermi level, a property inherited from the 2D bulk structure.\cite{Mastrogiuseppe2014}

\begin{figure}
  \centering
  \includegraphics[width=0.48\textwidth]{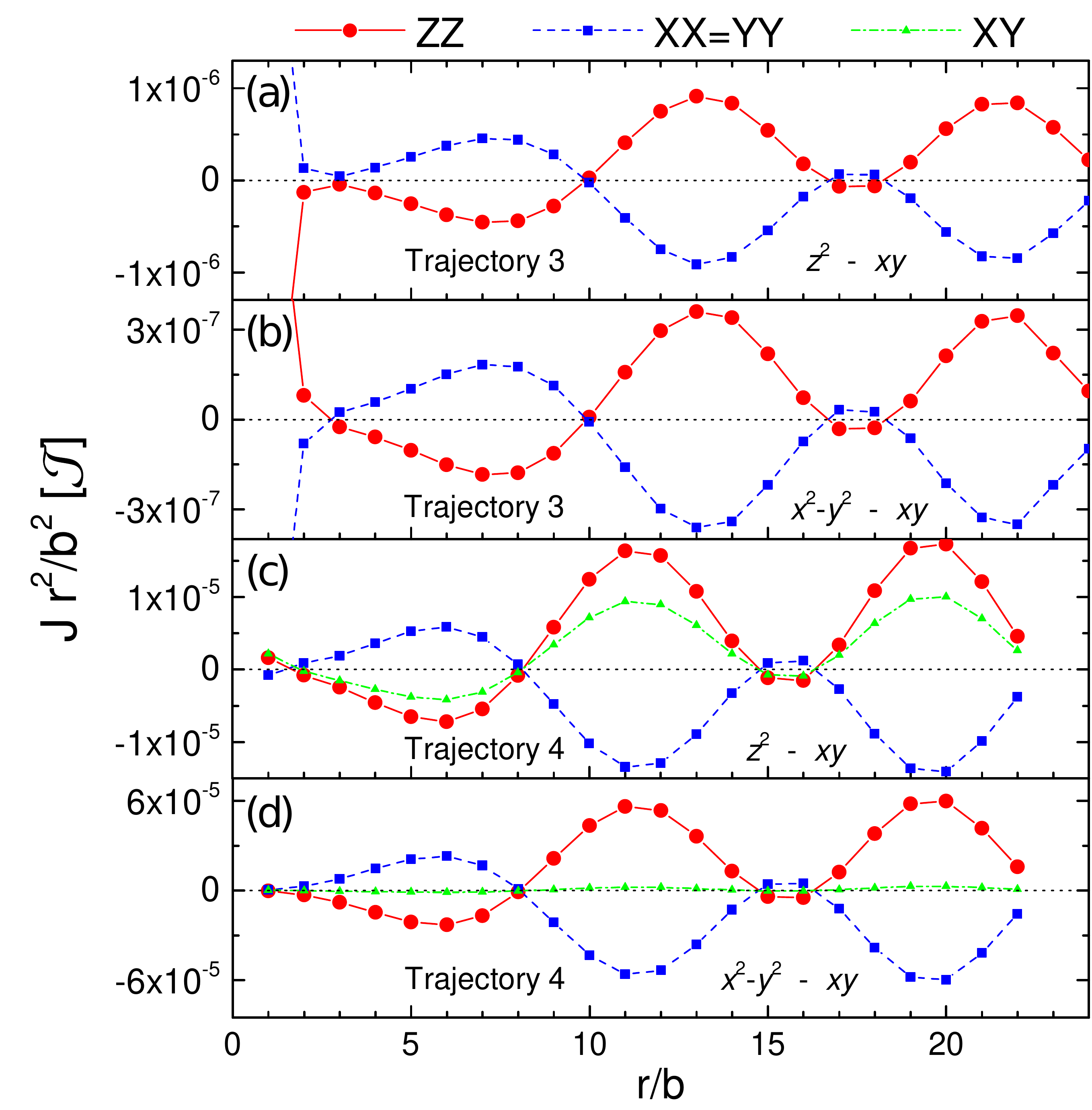}
  \caption{(Color online) The three components of the effective impurity interaction, scaled by $(r/b)^{2}$, versus relative distance along  armchair directions, with $b=a\sqrt{3}=|\bm{a}_{2}+\bm{a}_{3}|$. All curves correspond to $\varepsilon_{F1}$ and onsite hybridization to the orbitals indicated in the panels. Trajectories are explained in Fig.\ \ref{fig3}(a). Notice that in (a) and (b) there is no DM interaction, since the impurities lie on the line bisecting the flake, where reflection symmetry forbids its appearance.} \label{fig5}
\end{figure}

The interactions along \emph{armchair} directions are shown in Fig.\ \ref{fig5}(a-d). \emph{Trajectory} 3 follows a high symmetry line where the impurities lie on the line bisecting the flake [see Fig.\ \ref{fig3}(a)]. The interaction is modulated mostly by the large amplitude of $d_{xy}$ and $d_{x^2-y^2}$ orbitals, as shown in Fig.\ \ref{fig3}(c). Figure\ \ref{fig5}(a) shows $J^{z^2,xy}$, scaled by $(r/b)^{2}$, versus the relative distance between impurities in units of $b=a\sqrt{3}=|\bm{a}_{2}+\bm{a}_{3}|$, the nearest neighbor distance along armchair directions. The interaction is much weaker than the corresponding exchange along the zigzag directions. $J^{z^2,x^2-y^2}$ has very similar behavior. We can see that both $J_{ZZ}$ and $J_{XX}$ have a long-period oscillation, in contrast to the zigzag case, with period $8 b\approx 14 a$, and out of phase with each other. Most importantly, notice $J_{XY}=0$ for any orbital hybridization, reflecting the perfect cancellation seen in the infinite monolayer for impurities placed along the armchair direction.\cite{Mastrogiuseppe2014} Figure\ \ref{fig5}(b) shows $J^{x^2-y^2,xy}$ along the same trajectory. The interaction is of the same order of magnitude and shows the same behavior as $J^{z^2,xy}$, although slightly smaller in magnitude due to a suppressed $d_{x^2-y^2}$ at the bottom of the flake. We notice similar features as in Fig.\ \ref{fig5}(a), with an absence of DM interaction due to symmetry, and the long-period oscillation of the remaining components. To highlight the importance of symmetry, we now move the impurities along the armchair \emph{trajectory} 4, displaced laterally with respect to the vertical bisecting line of the triangle. The lack of reflection symmetry now allows the DM term to appear, although with smaller amplitude than the other component, as seen in  in Fig.\ \ref{fig5}(c) for $J^{z^2,xy}$. An even weaker DM interaction results for $J^{x^2-y^2,xy}$, as shown in Fig.\ \ref{fig5}(d). In all these interactions we see a long wavelength spatial modulation, signaling intravalley scattering processes.

\begin{figure}
	\centering
	\includegraphics[width=0.48\textwidth]{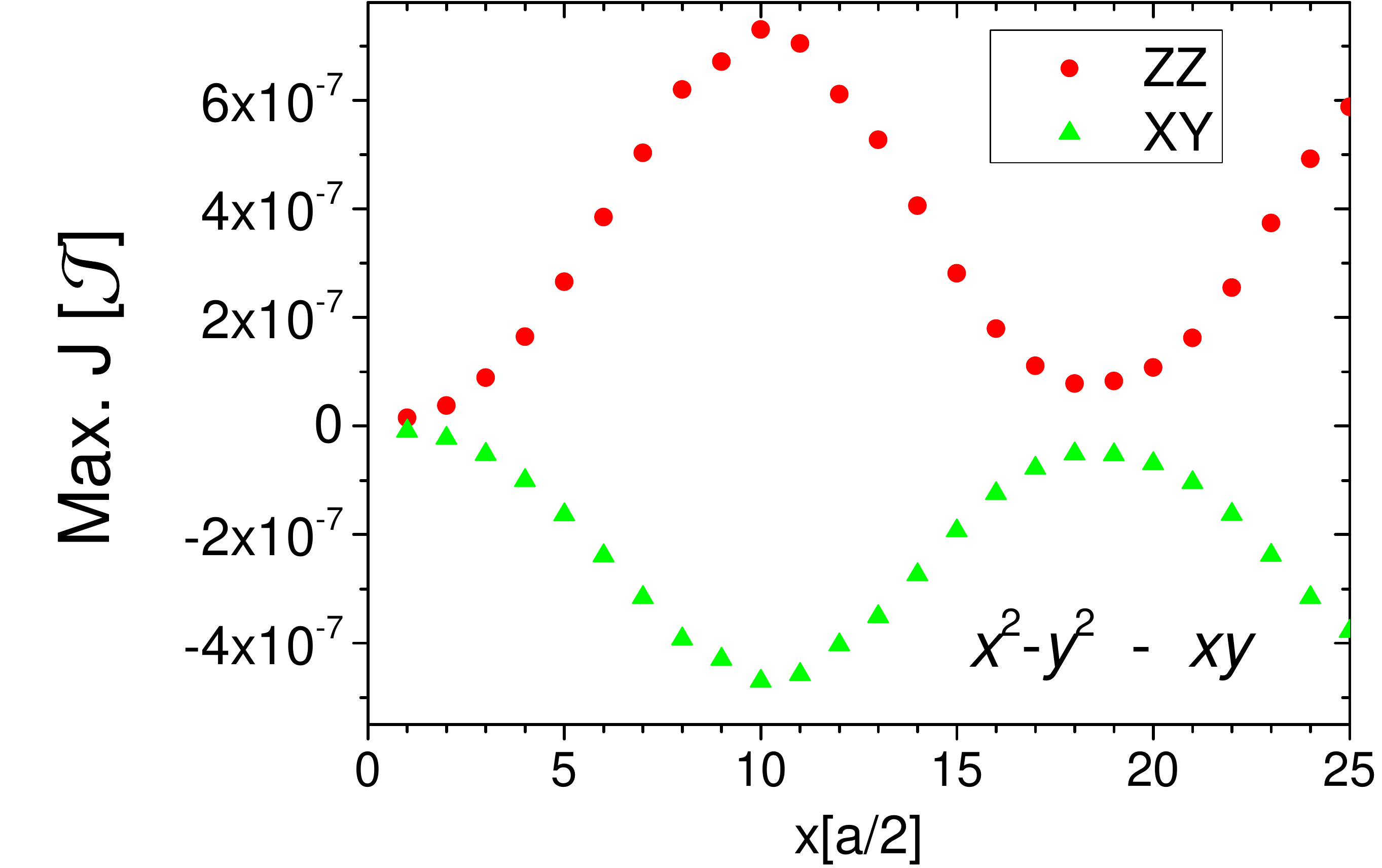}
	\caption{(Color online) Maximum interaction strength versus separation of parallel armchair trajectories \emph{with respect to the central bisecting line}.  Curves correspond to $\varepsilon_{F1}$ and onsite hybridization. The zero value in $x$ represents trajectory 3 and, as the vertical trajectory moves to the right, $x>0$, it approaches  the edge of the flake.}\label{fig6}
\end{figure}

In this finite triangular flake, $J_{XY}$ is always present for any zigzag trajectory, and for armchair trajectories along lines with lower symmetry. The only trajectory which respects reflection symmetry is indeed \emph{trajectory} 3. Displacing the armchair trajectory further away toward the edge of the flake results in larger $J_{XY}$, in general, although strongly modulated by the spatial dependence of the different orbital components of the states near the Fermi level. To illustrate this point, we follow the strength of two interaction terms, $J_{ZZ}$ and $J_{XY}$, for armchair trajectories that follow vertical lines parallel to the bisecting line of the flake. Figure\ \ref{fig6} shows the characteristic values of the interaction for orbitals $d_{x^2-y^2},d_{xy}$, as a function of the distance from the middle of the flake.
We track the maximum in each $J$ for impurity separations that lie in the interval $r/b\in [10, 15]$. The horizontal axis in Fig.\ \ref{fig6} indicates the $x$-distance from the bisecting line, where $x=0$ corresponds to \emph{trajectory} 3, and larger $x$ indicate armchair trajectories that are closer to the edge of the flake. We see that both $J_{ZZ}$ and $J_{XY}$ maintain their sign, either AFM or FM respectively, as the trajectories are displaced. The maxima are clearly modulated in both $J_{ZZ}$ and $J_{XY}$, reaching the largest amplitude at $x\simeq10\left(\frac{a}{2}\right)$, which is the characteristic length scale of the wave function antinode lobes in Fig.\ \ref{fig3}(c). The strong modulation of different interaction terms due to the wave function spatial patterns is ubiquitous in finite systems and provides another way to tune or find the most favorable or desired interaction between impurities. These results also highlight the importance of crystal symmetries in the interaction, further complicated by the shape of the finite flake, as diverse as stars \cite{Van2013}, hexagons\cite{Cao2015hexagonalflakes}, and rhomboids\cite{Wang2013rhomboidflakes}, among others, in experimental systems.

\subsubsection{Doping level $\varepsilon_{F2}$}

For a deeper doping level, such as $\varepsilon_{F2}$ [Fig.\ \ref{fig2}(b)], the bulk monolayer has contributions from the bands at the $\Gamma$ point, which introduces $\Gamma$-$K$($K'$) intervalley scattering. The  magnitude squared of the wave functions for this level are shown in Fig.\ \ref{fig3}(d)-(f). In this case, the wave function is dominated by the $d_{z^2}$ component, as one would expect from the strong $\Gamma$ content. As the Fermi energy gets deeper into the valence band, the states are also more bulk-like, extending throughout the crystal flake with all three orbital components.

We find that the indirect exchange in zigzag \emph{trajectories} 1 and 2 has similar behavior to the one described for $\varepsilon_{F1}$, with natural quantitative differences on the overall amplitude, which turns out to be two or three orders of magnitude larger, depending on the orbital to which impurities  hybridize, and on the spatial modulation of the wave functions near the Fermi level. The interactions (not shown) oscillate between FM and AFM behavior, with additional frequencies and modulations, reflecting the participation of energy states from the spin-degenerate band at the $\Gamma$ point, which provides a sizable contribution to the scattering processes. The interplay between different valleys and subtle wave function modulations result in a complex oscillatory pattern for the different exchange components. We observe larger strength, the appearance of beatings, and subtle interaction modulations as the different scattering processes compete with each other. This is very similar to the behavior seen in 2D bulk systems at these doping levels \cite{Mastrogiuseppe2014}, with strong noncolinear interaction $J_{XY}$, as well as $J_{ZZ}$ and $J_{XX}$, which adds to the tunability and complexity of the resulting interaction.
%
\begin{figure}
  \centering
  \includegraphics[width=0.48\textwidth]{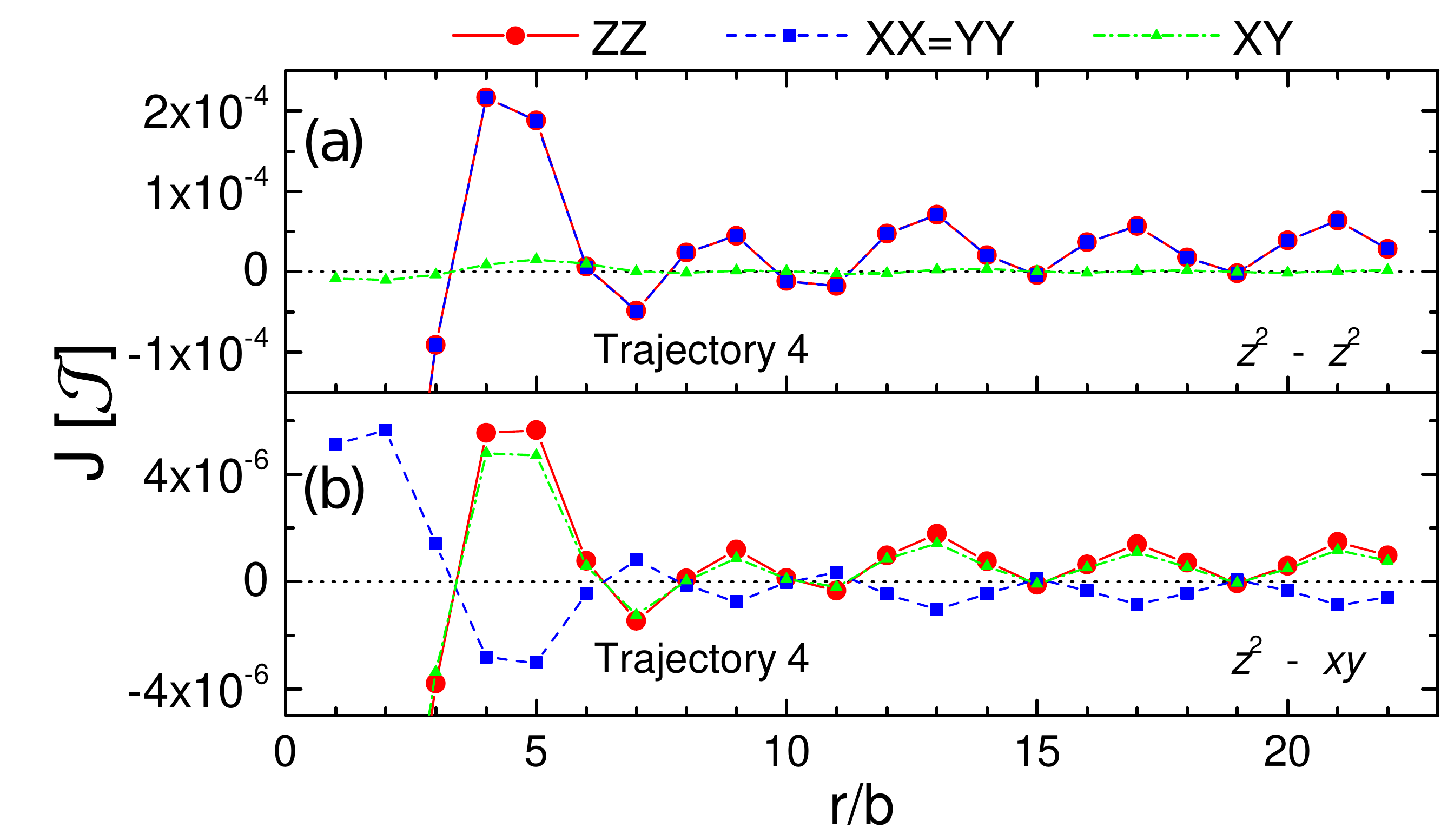}
  \caption{(Color online) Effective interaction versus relative distance along armchair direction. These results correspond to $\varepsilon_{F2}$ and onsite hybridization.  (a) When both impurities hybridize to $d_{z^2}$ orbitals, a Heisenberg-like interaction is found. (b) For this pair of orbitals, XX gets out of phase with coinciding Ising and DM terms.}\label{fig7}
\end{figure}
We find somewhat different behavior for exchange interactions along armchair directions. The results are shown in Fig.\ \ref{fig7}, with $J^{z^2,z^2}$ and $J^{z^2,xy}$ along \emph{trajectory} 4. It is interesting that the interaction decays much more slowly than $1/r$, signaling the strong size quantization of the $d_{z^2}$ component, which dominates these interactions. Notice in Fig.\ \ref{fig7}(a) that the Ising and XX terms match (the same as in the zigzag case for this doping). As the DM interaction is vanishingly small, the net interaction is Heisenberg-like: collinear and symmetric. On the other hand, Fig.\ \ref{fig7}(b) shows that $J_{ZZ}$ and $J_{XY}$ are nearly in phase with each other, competing against $J_{XX}$, which turns out to be out of phase with the previous two. Notice as well that for this $d_{xy}$ hybridization, the amplitude of the interactions is largely suppressed.

\subsubsection{Varying doping levels}

\begin{figure}
	\includegraphics[width=0.5\textwidth]{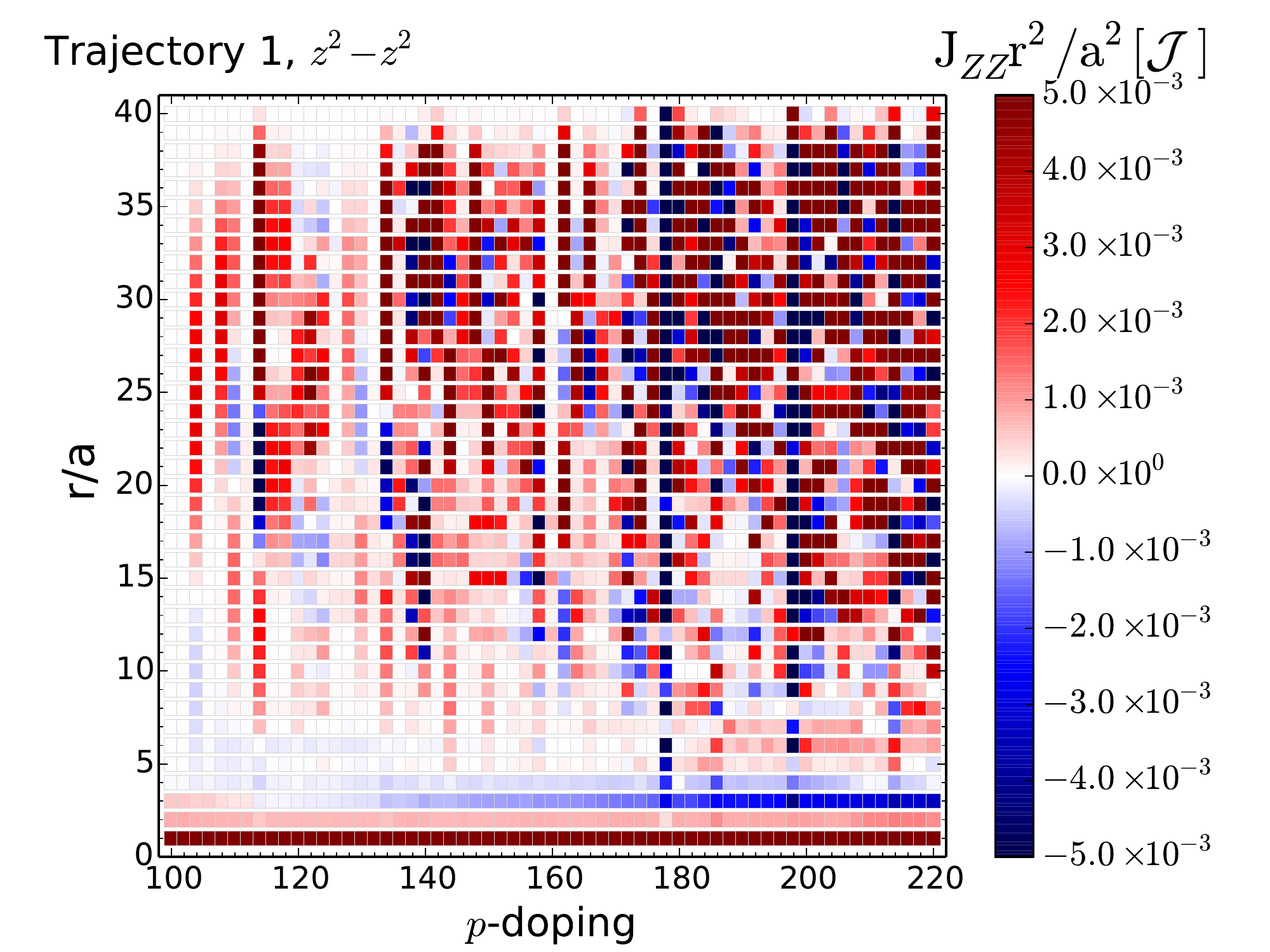}
	\caption{(Color online) Ising component of the indirect exchange, scaled by $(r/a)^{2}$, for various levels of $p$-doping, with trajectory and orbital hybridization as indicated in the figure. Positive (negative) values correspond to AFM (FM) impurity alignment.}\label{fig8}
\end{figure}

Figure\ \ref{fig8} represents a two-dimensional map of the Ising component of the indirect exchange [scaled by $(r/a)^{2}$], as a function of the $p$-doping, represented by the number of holes in the sample. Both impurities are hybridized to $d_{z^2}$ orbitals and displaced along \emph{trajectory} 1.
One can observe that for some doping levels the interaction is always FM or AFM, and for others it changes sign along the trajectory. Notice that, in general, as one gets deeper into the valence band, the magnitude of the interaction increases. This is an expected behavior because, as the Fermi level decreases, the energy states get more densely packed, providing more access to low-energy particle-hole excitations.
In conclusion, the control of the doping level provides an interesting tunability tool for the indirect exchange.

\subsection{Plaquette Hybridization}
\label{subsec:plaquette}

\begin{figure*}
	\centering
	\includegraphics[width=0.48\textwidth]{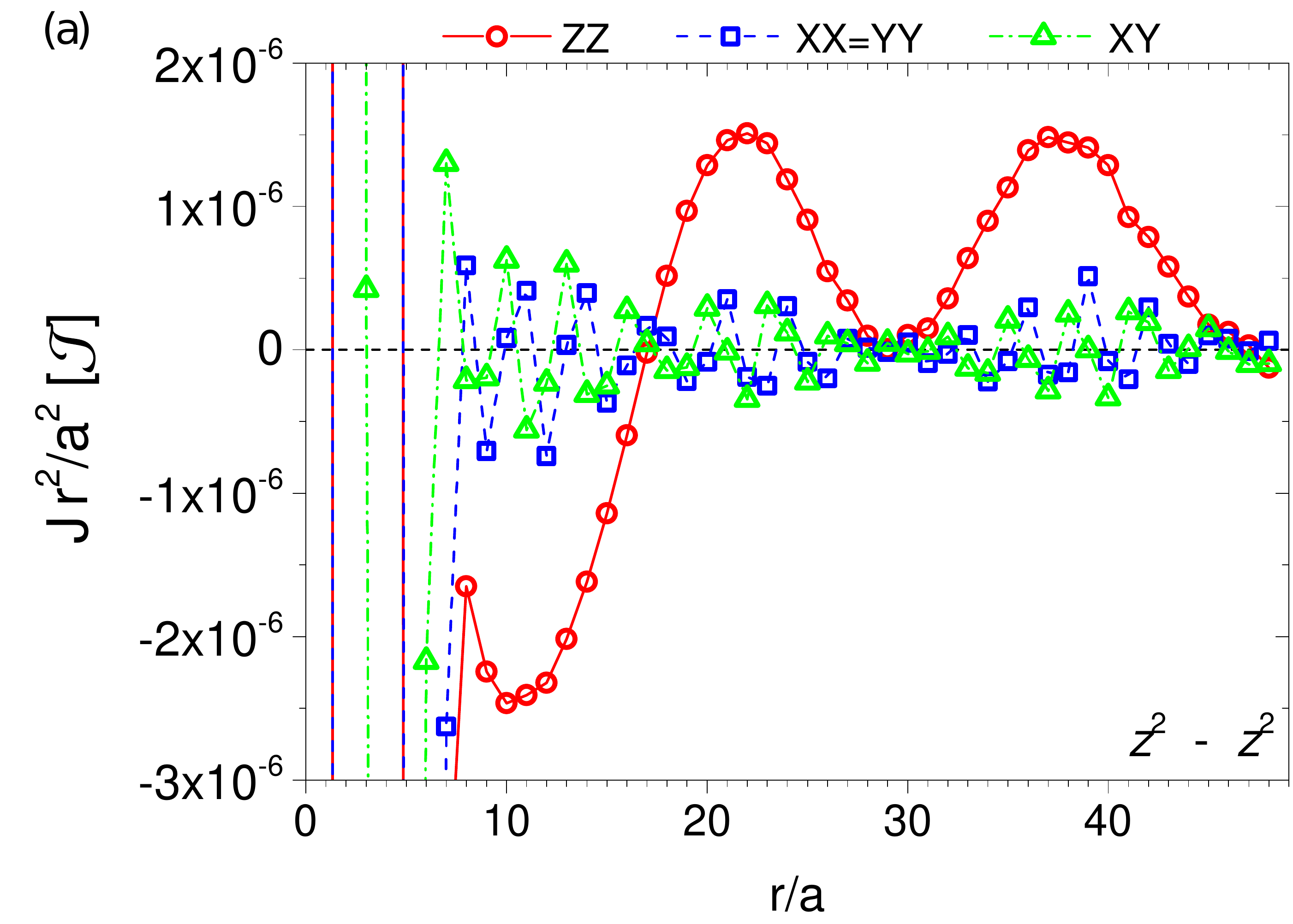}
	\includegraphics[width=0.48\textwidth]{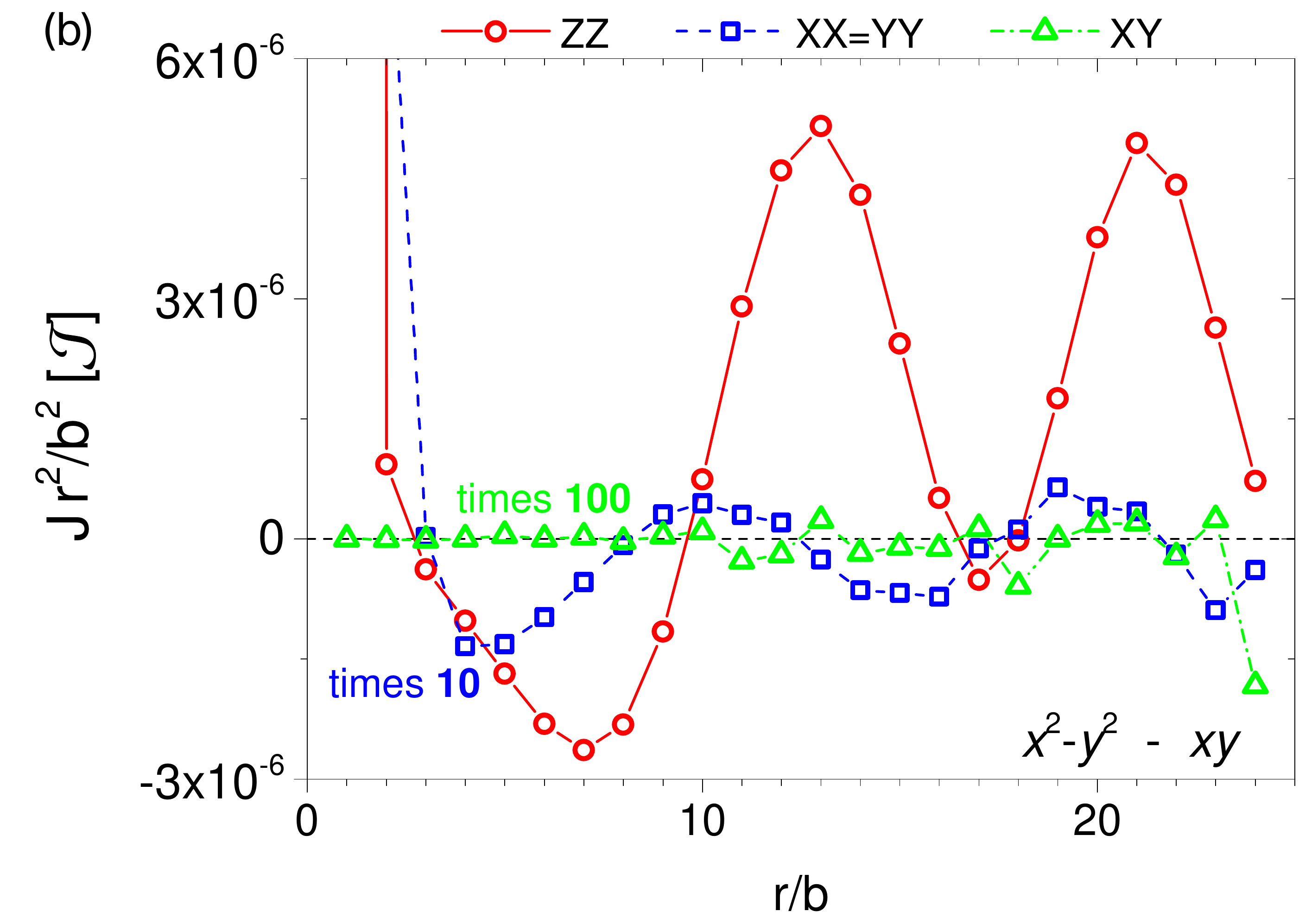}
	\caption{(Color online) (a) Effective impurity interaction, scaled by $(r/a)^{2}$, versus the relative distance in zigzag direction between the impurities $r$. All curves for $\varepsilon_{F1}$ and plaquette triangle \emph{down} configuration. The first magnetic impurity is located at bottom corner of the flake (surrounded by sites 1, 2 and 3), and the second one moves along the zigzag edge on the right. (b) The same, but for armchair trajectory 3 (notice that the scales for $XX=YY$ and $XY$ have been amplified 10 and 100 times respectively, for better visualization). Orbitals are indicated in each panel.}\label{fig9}
\end{figure*}

We now study the role of different atomic environments on the effective exchange interaction, focusing on ``plaquette" or ``hollow" sites. This kind of impurity environment has been found stable for Fe and Mn adatoms, and associated to either adatoms on a pristine monolayer or on disulfur vacancies\cite{Cong2015}. These environments are associated with two different hollow sites with three-fold symmetry, which one can identify as triangle \emph{up} and triangle \emph{down} environments. Figures\ \ref{fig1}(b) and (c) show plaquette impurities in a triangle \emph{down} configuration, which in the lattice correspond to hollow sites in hexagons formed by Mo and S$_{2}$ atoms. In triangle \emph{up} configurations (not shown), the impurities sit on a disulfur location, also equidistant from the three Mo atoms.
In either case, the RKKY interaction is composed of an interference of 9 scattering terms, corresponding to a combination of onsite interactions between pairs of atoms that surround each impurity.
For instance, if $|\bm r|$ denotes the distance between the lower vertices of each triangle (in the triangle down environment), then we have 3 interactions with distance $|\bm r|$, and the remaining six correspond to distances given by $|\bm r \pm \bm a_l|$, with $l=1,2,3$.
%
Let us study the plaquette triangle \emph{down}  configuration, with  impurities following zigzag and armchair trajectories. For the zigzag case, the first impurity is fixed at the lower corner of the sample. For armchair, we study \emph{trajectory} 3. The doping level is set to $\varepsilon_{F1}$. Each impurity hybridizes to three surrounding Mo atoms, with an exchange coupling of $\frac{{\cal J}}{3}$ to each of them. In Fig.\ \ref{fig9}, we show the spatial dependence of the indirect interaction $J^{z^2,z^2}$ for the zigzag and $J^{x^2-y^2,xy}$ for the armchair direction respectively,. We observe the typical quadratic decay, and also fast and slow oscillations for the in-plane $J_{XX}$, $J_{XY}$ and Ising $J_{ZZ}$ terms, respectively, in the zigzag direction. In the armchair direction, notably, the in-plane components are strongly reduced in magnitude.
Although the previous features agree with the ones seen for the onsite configuration, there is a notable difference. In the plaquette case, the Ising $J_{ZZ}$ interaction term has larger magnitude than the in-plane  ones, as one can see in Fig.\ \ref{fig9}. If we compare the zigzag cases, we observe that $J_{ZZ}$ detaches from the envelope of the modulation created by $J_{XX}$ and $J_{XY}$ by a typical factor of 2 or 3 times larger in magnitude.
In the armchair direction, the detaching is more dramatic, as seen in Fig.\ \ref{fig9}. As in Fig.\ \ref{fig4}, the intra- (for $J_{ZZ}$) and inter-valley (for $J_{XX}$ and $J_{XY}$) processes are the scattering mechanisms responsible for the interaction wavelengths.

\begin{figure*}
	\centering
	\includegraphics[width=0.48\textwidth]{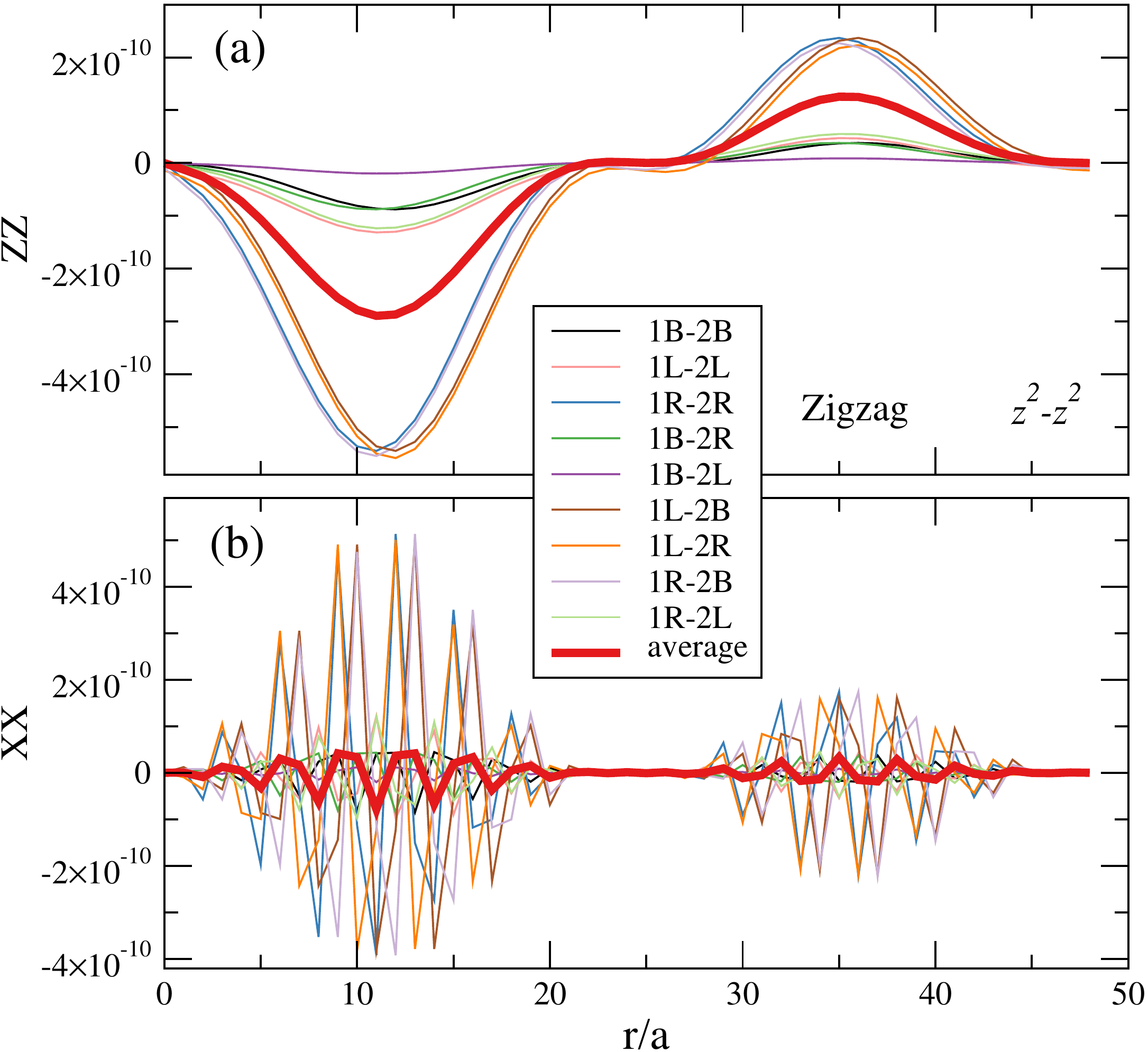}
	\includegraphics[width=0.459\textwidth]{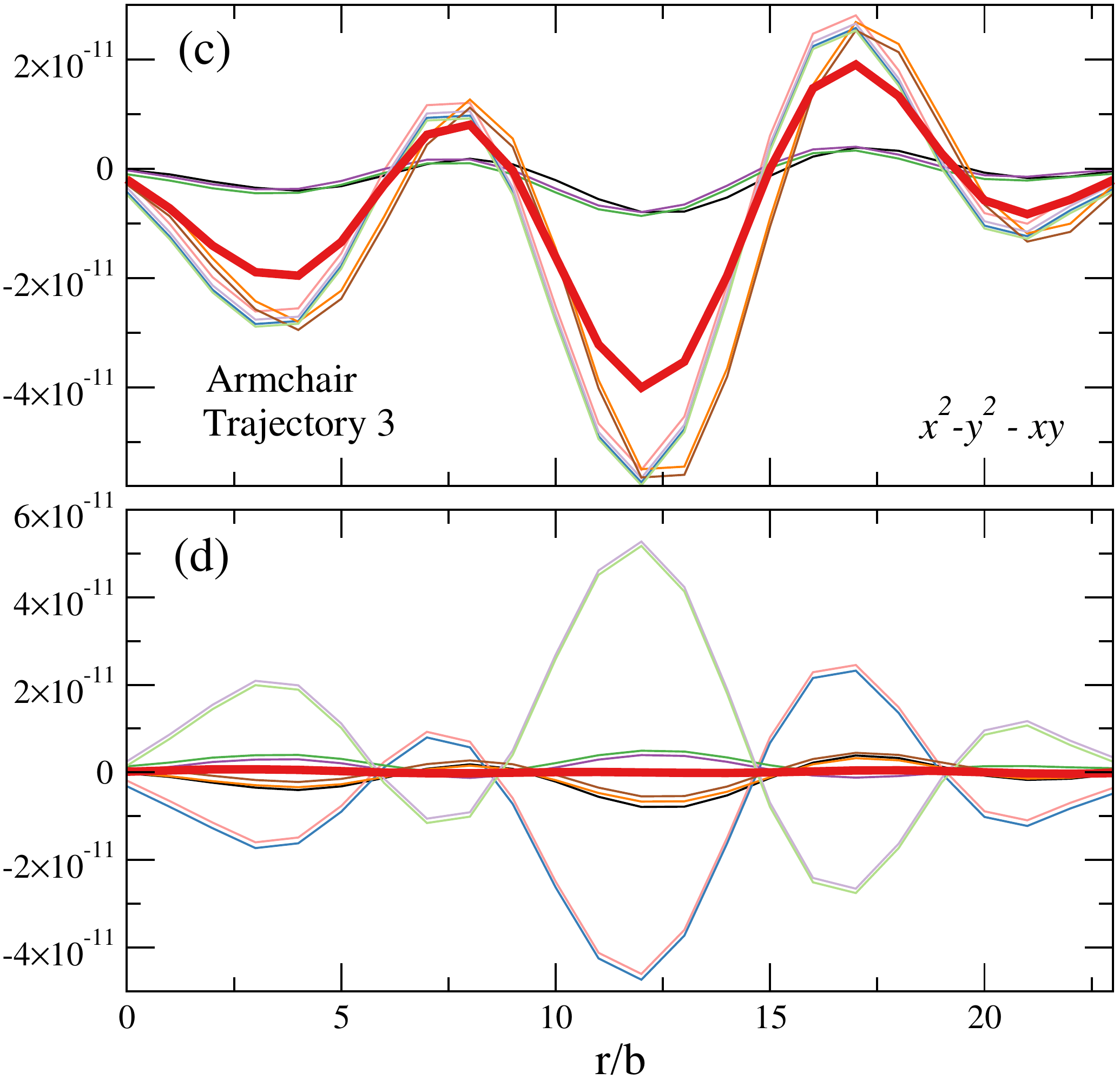}
	\caption{(Color online) Contribution of the two lowest energy particle-hole excitations, obtained from perturbation theory, to the effective impurity exchange components \emph{ZZ} and \emph{XX}, in the triangle down plaquette configuration. Trajectories and orbitals are the same as in Fig.\ \ref{fig9}. The legend for each thin curve, $1\gamma-2\gamma'$, with $\gamma,\gamma' \in \{$B, L, R$\}$, indicate the first (1) and second impurity (2) connection to the $\{$Bottom, Left, Right$\}$ Mo atom in the respective surrounding triangle. The thick curve indicates the average of the 9 onsite terms.}\label{fig10}
\end{figure*}

To gain understanding of this behavior, we analyze the terms corresponding to the lowest two particle-hole excitations in perturbation theory (see Appendix \ref{app:pertub} for calculation details).
Figures\ \ref{fig10}(a) and (b) show the most relevant components of the $J_{ZZ}$ and $J_{XX}$ interaction terms in the zigzag direction defined with the first impurity at the hollow site of the triangle in the bottom corner of the flake, and both impurities hybridized to $d_{z^2}$ orbitals.
Each panel shows curves for the 9 different onsite interaction terms, together with the average. As one can observe, for $ZZ$ all the long-wavelength components are in phase, resulting in an average of the same order of the individual onsite components. On the other hand, for $XX$, we can see that the short-wavelength components  get out of phase, resulting in a suppressed average interaction. The case is similar for the $XY$ term. This would explain the detaching behavior seen in Fig.\ \ref{fig9}(a).
In the armchair trajectory, a similar situation occurs, as seen in Fig.\ \ref{fig10}(c) and (d).
Again, $ZZ$ has all its onsite components in phase, whereas $XX$ has out of phase components that almost perfectly cancel each other, resulting in a negligible $XX$ term. This detaching behavior is not seen for any case in the onsite configuration, and provides an extra tunable tool when the impurities are hybridized in a plaquette environment.
Notice that the perturbation results of Fig.\ \ref{fig10} provide just a qualitative explanation of the real picture, because only the two lowest particle-hole excitations are shown. By adding up higher energy processes, the oscillations start looking like the ones in Fig.\ \ref{fig9}.

The results for the plaquette triangle \emph{up} absorption configuration are similar to the ones of triangle \emph{down} (off by a typical magnitude factor of $1/10$ and shifted by one $r/a$ period). As discussed before, the interaction is largely influenced by the wave function modulation, and adjacent \emph{up} and \emph{down} triangles do not possess the same wave function distribution, although it is quite similar.

\section{Discussion}
\label{sec:conclusions}

In this paper, we investigated the effective indirect interaction between two magnetic impurities embedded in a \emph{p}-doped triangular zigzag-terminated MoS$_{2}$ flake. We analyzed the interaction when impurities are displaced along various trajectories, including bulk and edge cases, and considering hybridization to different transition metal orbitals.
We studied onsite and plaquette configurations, which are the most probable adsorption sites from an experimental point of view. We concentrated on two levels of hole doping, and also provided an example of the interaction as a function of the impurity separation, for a range of doping levels.

As a general rule, the interaction decays as $r^{-2}$, as in conventional 2D electron gases. However, there can be exceptions for which the decay is slower.
The interactions show long wavelength spatial modulations along armchair directions, and for the Ising component along zigzag directions, signaling intravalley scattering processes which conserve the spin projection. On the other hand, the in-plane components along zigzag directions display short-period oscillations, signaling intravalley scattering processes that flip the spin.

We have also found that the symmetries of the host play an important role in determining the behavior of the interaction. In the infinite MoS$_2$ monolayer, it was predicted that the DM interaction vanishes along the armchair direction due to lattice reflection symmetry \cite{Mastrogiuseppe2014}. Here, we showed that this property holds only when considering a trajectory along the vertical line bisecting the triangular flake, which is the only direction that respects this symmetry.

For the triangle-down plaquette configuration, we found that the Ising interaction is larger than the in-plane ones. We provided a qualitative explanation of this phenomenon, calculating two components of the interaction, corresponding to the lowest particle-hole excitations in perturbation theory. For the Ising component, each of the 9 individual onsite terms associated with scattering processes between pairs of atoms surrounding each impurity,  turn out to be in phase, giving a constructive interference that results in a sizable average value of $J_{ZZ}$. For the in-plane interactions, different components turn out to be out of phase, producing a reduced plaquette interaction.

At given doping levels, the distribution of the modulus of the wave function on the sample can be used as a qualitative guide to tune the strength of the RKKY interaction. In particular, an scanning tunneling spectroscopy (STS) experiment over TMD flakes could be used to
map the local density of states (LDOS) over the sample, and use the microscope tip to embed magnetic impurities in regions with high LDOS \cite{Nipane2016}. A spin polarized tip can then measure the resulting indirect exchange.
All in all, our results provide tools for designing noncolinear arrangements between impurities, suggesting interesting long range ordering of spin chains and 2D arrays of magnetic moments in these materials.


\begin{acknowledgments}
We acknowledge support from NSF-DMR 1508325. O. \'A.-O. acknowledges a research fellowship from the Condensed Matter and Surface Science program at Ohio University. We thank Don Roth for help in implementing the computational calculations in our cluster.
\end{acknowledgments}

\appendix

\section{Perturbation theory}\label{app:pertub}

The effective exchange integrals can also be calculated in perturbation theory \cite{Mattis, Nolting}, for small ${\cal J}_{\alpha_i}$ in Eq.\ \eqref{impurities1}.
Considering for simplicity that the local hybridization parameter between conduction electrons and impurities $\mathcal{J}$ is the same for every orbital, we can rewrite Eq.\ \eqref{impurities1} as
\begin{equation}\label{impurities1_2}
H_{\text{I}}= \mathcal{J} \sum_{i=1,2}  S_{i}^z s^z_{\alpha_i}(\bm{r}_i) + \frac12\left[S_{i}^+ s^-_{\alpha_i}(\bm{r}_i) + S_{i}^- s^+_{\alpha_i}(\bm{r}_i)\right],
\end{equation}
with
\begin{align}\label{impurities2_2}
\begin{split}
s^z_{\alpha}(\bm{r}_j)&=\frac{1}{2}\left[d_{\alpha,\ua}^{\dagger}(\bm{r}_j) d_{\alpha,\ua}(\bm{r}_j) - d_{\alpha,\da}^{\dagger}(\bm{r}_j) d_{\alpha,\da}(\bm{r}_j)\right],\\
s^+_{\alpha}(\bm{r}_j)&=d_{\alpha,\ua}^{\dagger}(\bm{r}_j) d_{\alpha,\da}(\bm{r}_j),\\
s^-_{\alpha}(\bm{r}_j)&=d_{\alpha,\da}^{\dagger}(\bm{r}_j) d_{\alpha,\ua}(\bm{r}_j).
\end{split}
\end{align}

In the basis that diagonalizes $H_0$, defined in Eq.\ \eqref{eq:diagonaliz}, the spin operators read
\begin{align}\label{eq:spin_diag}
\begin{split}
s^z_{\alpha}(\bm{r}_j)&=\frac{1}{2}\sum_{\mu,\mu'}\left[\psi_{k,\mu}^* \psi_{k,\mu'} c_{\mu,\ua}^{\dagger} c_{\mu',\ua} - \psi_{k,\mu} \psi_{k,\mu'}^* c_{\mu,\da}^{\dagger} c_{\mu',\da}\right],\\
s^+_{\alpha}(\bm{r}_j)&=\sum_{\mu,\mu'} \psi_{k,\mu}^* \psi_{k,\mu'}^* c_{\mu,\ua}^{\dagger} c_{\mu',\da},\\
s^-_{\alpha}(\bm{r}_j)&=\sum_{\mu,\mu'} \psi_{k,\mu} \psi_{k,\mu'} c_{\mu,\da}^{\dagger} c_{\mu',\ua}.
\end{split}
\end{align}

The  second order correction to the energy in perturbation theory is given by
\begin{equation}\label{eq:perturb}
E^{(2)} = \sum_{ex,\mathcal{D}'}\frac{|\Braket{GS;\mathcal{D}|H_I|ex;\mathcal{D}'}|^2} {E_{GS} - E_{ex}}.
\end{equation}
In this expression, $\ket{GS;\mathcal{D}}\equiv \ket{GS}\ket{\mathcal{D}}$, where $\ket{GS}$ is the ground state of $H_0$ and $\ket{\mathcal{D}}$ the ground state spin configuration of the two disconnected magnetic moments. Similarly, $\ket{ex}$ denote particle-hole excitations of the electron gas, and $\ket{\mathcal{D}'}$ are excited configurations of the two impurities.
Inserting \eqref{eq:spin_diag} in \eqref{impurities1_2}, one can compute expression \eqref{eq:perturb}. After some algebra, one gets
\begin{equation}
\begin{split}
E^{(2)}_{\alpha,\alpha'} = \frac{\mathcal{J}^2}{2}&\sum_{\substack{\mu \leq \mu_F\\\mu'>\mu_F}}\frac{1}{\epsilon_\mu -\epsilon_{\mu'}}\bra{\mathcal{D}}
J_{ZZ}^{\alpha,\alpha'}(\bm{r}_j,\mathbf{r}_{j'})  S_j^z S_{j'}^z \\
&+ J_{XX}^{\alpha,\alpha'}(\bm{r}_j,\mathbf{r}_{j'})  (S_j^x S_{j'}^x + S_j^y S_{j'}^y) \\
&+ J_{XY}^{\alpha,\alpha'}(\bm{r}_j,\mathbf{r}_{j'})  (S_j^x S_{j'}^y - S_j^y S_{j'}^x)\ket{\mathcal{D}}
\end{split}
\end{equation}
with
\begin{align}
\begin{split}
J_{ZZ}^{\alpha,\alpha'} (\bm{r}_j,\mathbf{r}_{j'})& = \sum_{\substack{\mu \leq \mu_F\\\mu'>\mu_F}}\text{Re} \left[ (\psi_{j, \mu}^\alpha)^* \psi_{j, \mu'}^\alpha \psi_{j', \mu}^{\alpha'} (\psi_{j', \mu'}^{\alpha'})^*\right],\\
J_{XX}^{\alpha,\alpha'}(\bm{r}_j,\mathbf{r}_{j'})& = \sum_{\substack{\mu \leq \mu_F\\\mu'>\mu_F}}\text{Re} \left[ \psi_{j, \mu}^\alpha \psi_{j, \mu'}^\alpha (\psi_{j', \mu}^{\alpha'})^* (\psi_{j', \mu'}^{\alpha'})^*\right],\\
J_{XY}^{\alpha,\alpha'}(\bm{r}_j,\mathbf{r}_{j'})& = -\sum_{\substack{\mu \leq \mu_F\\\mu'>\mu_F}}\text{Im} \left[ \psi_{j, \mu}^\alpha \psi_{j, \mu'}^\alpha (\psi_{j', \mu}^{\alpha'})^* (\psi_{j', \mu'}^{\alpha'})^*\right].
\end{split}
\end{align}
In these last expressions, we have used the short-hand notation for the eigenvectors introduced in the main text. $\mu_F$ denotes the level index associated with a given Fermi energy $\varepsilon_{F}$ in the TMD flake. The curves shown in Fig.\ \ref{fig10} correspond to $\mu=\mu_F$,  $\mu'=\mu_F+1$, and $\mu'=\mu_F+2$, where $\mu_F$ correspond to 106 holes or, equivalently, $\varepsilon_{F1}$ in the main text.
\vspace{0.1cm}

\bibliographystyle{apsrev4-1}
\bibliography{biblio}

\begin{thebibliography}{69}%
\makeatletter
\providecommand \@ifxundefined [1]{%
 \@ifx{#1\undefined}
}%
\providecommand \@ifnum [1]{%
 \ifnum #1\expandafter \@firstoftwo
 \else \expandafter \@secondoftwo
 \fi
}%
\providecommand \@ifx [1]{%
 \ifx #1\expandafter \@firstoftwo
 \else \expandafter \@secondoftwo
 \fi
}%
\providecommand \natexlab [1]{#1}%
\providecommand \enquote  [1]{``#1''}%
\providecommand \bibnamefont  [1]{#1}%
\providecommand \bibfnamefont [1]{#1}%
\providecommand \citenamefont [1]{#1}%
\providecommand \href@noop [0]{\@secondoftwo}%
\providecommand \href [0]{\begingroup \@sanitize@url \@href}%
\providecommand \@href[1]{\@@startlink{#1}\@@href}%
\providecommand \@@href[1]{\endgroup#1\@@endlink}%
\providecommand \@sanitize@url [0]{\catcode `\\12\catcode `\$12\catcode
  `\&12\catcode `\#12\catcode `\^12\catcode `\_12\catcode `\%12\relax}%
\providecommand \@@startlink[1]{}%
\providecommand \@@endlink[0]{}%
\providecommand \url  [0]{\begingroup\@sanitize@url \@url }%
\providecommand \@url [1]{\endgroup\@href {#1}{\urlprefix }}%
\providecommand \urlprefix  [0]{URL }%
\providecommand \Eprint [0]{\href }%
\providecommand \doibase [0]{http://dx.doi.org/}%
\providecommand \selectlanguage [0]{\@gobble}%
\providecommand \bibinfo  [0]{\@secondoftwo}%
\providecommand \bibfield  [0]{\@secondoftwo}%
\providecommand \translation [1]{[#1]}%
\providecommand \BibitemOpen [0]{}%
\providecommand \bibitemStop [0]{}%
\providecommand \bibitemNoStop [0]{.\EOS\space}%
\providecommand \EOS [0]{\spacefactor3000\relax}%
\providecommand \BibitemShut  [1]{\csname bibitem#1\endcsname}%
\let\auto@bib@innerbib\@empty
\bibitem [{\citenamefont {Novoselov}\ \emph {et~al.}(2005)\citenamefont
  {Novoselov}, \citenamefont {Jiang}, \citenamefont {Schedin}, \citenamefont
  {Booth}, \citenamefont {Khotkevich}, \citenamefont {Morozov},\ and\
  \citenamefont {Geim}}]{Novoselov2005}%
  \BibitemOpen
  \bibfield  {author} {\bibinfo {author} {\bibfnamefont {K.~S.}\ \bibnamefont
  {Novoselov}}, \bibinfo {author} {\bibfnamefont {D.}~\bibnamefont {Jiang}},
  \bibinfo {author} {\bibfnamefont {F.}~\bibnamefont {Schedin}}, \bibinfo
  {author} {\bibfnamefont {T.~J.}\ \bibnamefont {Booth}}, \bibinfo {author}
  {\bibfnamefont {V.~V.}\ \bibnamefont {Khotkevich}}, \bibinfo {author}
  {\bibfnamefont {S.~V.}\ \bibnamefont {Morozov}}, \ and\ \bibinfo {author}
  {\bibfnamefont {A.~K.}\ \bibnamefont {Geim}},\ }\href {\doibase
  10.1073/pnas.0502848102} {\bibfield  {journal} {\bibinfo  {journal} {Proc.
  Natl. Acad. Sci. U.S.A.}\ }\textbf {\bibinfo {volume} {102}},\ \bibinfo
  {pages} {10451} (\bibinfo {year} {2005})}\BibitemShut {NoStop}%
\bibitem [{\citenamefont {Geim}\ and\ \citenamefont
  {Grigorieva}(2013)}]{Geim2013}%
  \BibitemOpen
  \bibfield  {author} {\bibinfo {author} {\bibfnamefont {A.~K.}\ \bibnamefont
  {Geim}}\ and\ \bibinfo {author} {\bibfnamefont {I.~V.}\ \bibnamefont
  {Grigorieva}},\ }\href {\doibase 10.1038/nature12385} {\bibfield  {journal}
  {\bibinfo  {journal} {Nature}\ }\textbf {\bibinfo {volume} {499}},\ \bibinfo
  {pages} {419} (\bibinfo {year} {2013})}\BibitemShut {NoStop}%
\bibitem [{\citenamefont {Wang}\ \emph {et~al.}(2012)\citenamefont {Wang},
  \citenamefont {Kalantar-Zadeh}, \citenamefont {Kis}, \citenamefont
  {Coleman},\ and\ \citenamefont {Strano}}]{Wang2012}%
  \BibitemOpen
  \bibfield  {author} {\bibinfo {author} {\bibfnamefont {Q.~H.}\ \bibnamefont
  {Wang}}, \bibinfo {author} {\bibfnamefont {K.}~\bibnamefont
  {Kalantar-Zadeh}}, \bibinfo {author} {\bibfnamefont {A.}~\bibnamefont {Kis}},
  \bibinfo {author} {\bibfnamefont {J.~N.}\ \bibnamefont {Coleman}}, \ and\
  \bibinfo {author} {\bibfnamefont {M.~S.}\ \bibnamefont {Strano}},\ }\href
  {\doibase 10.1038/nnano.2012.193} {\bibfield  {journal} {\bibinfo  {journal}
  {Nat. Nanotechnol.}\ }\textbf {\bibinfo {volume} {7}},\ \bibinfo {pages}
  {699} (\bibinfo {year} {2012})}\BibitemShut {NoStop}%
\bibitem [{\citenamefont {Bhimanapati}\ \emph {et~al.}(2015)\citenamefont
  {Bhimanapati}, \citenamefont {Lin}, \citenamefont {Meunier}, \citenamefont
  {Jung}, \citenamefont {Cha}, \citenamefont {Das}, \citenamefont {Xiao},
  \citenamefont {Son}, \citenamefont {Strano}, \citenamefont {Cooper},
  \citenamefont {Liang}, \citenamefont {Louie}, \citenamefont {Ringe},
  \citenamefont {Zhou}, \citenamefont {Kim}, \citenamefont {Naik},
  \citenamefont {Sumpter}, \citenamefont {Terrones}, \citenamefont {Xia},
  \citenamefont {Wang}, \citenamefont {Zhu}, \citenamefont {Akinwande},
  \citenamefont {Alem}, \citenamefont {Schuller}, \citenamefont {Schaak},
  \citenamefont {Terrones},\ and\ \citenamefont {Robinson}}]{Bhimanapati2015}%
  \BibitemOpen
  \bibfield  {author} {\bibinfo {author} {\bibfnamefont {G.~R.}\ \bibnamefont
  {Bhimanapati}}, \bibinfo {author} {\bibfnamefont {Z.}~\bibnamefont {Lin}},
  \bibinfo {author} {\bibfnamefont {V.}~\bibnamefont {Meunier}}, \bibinfo
  {author} {\bibfnamefont {Y.}~\bibnamefont {Jung}}, \bibinfo {author}
  {\bibfnamefont {J.}~\bibnamefont {Cha}}, \bibinfo {author} {\bibfnamefont
  {S.}~\bibnamefont {Das}}, \bibinfo {author} {\bibfnamefont {D.}~\bibnamefont
  {Xiao}}, \bibinfo {author} {\bibfnamefont {Y.}~\bibnamefont {Son}}, \bibinfo
  {author} {\bibfnamefont {M.~S.}\ \bibnamefont {Strano}}, \bibinfo {author}
  {\bibfnamefont {V.~R.}\ \bibnamefont {Cooper}}, \bibinfo {author}
  {\bibfnamefont {L.}~\bibnamefont {Liang}}, \bibinfo {author} {\bibfnamefont
  {S.~G.}\ \bibnamefont {Louie}}, \bibinfo {author} {\bibfnamefont
  {E.}~\bibnamefont {Ringe}}, \bibinfo {author} {\bibfnamefont
  {W.}~\bibnamefont {Zhou}}, \bibinfo {author} {\bibfnamefont {S.~S.}\
  \bibnamefont {Kim}}, \bibinfo {author} {\bibfnamefont {R.~R.}\ \bibnamefont
  {Naik}}, \bibinfo {author} {\bibfnamefont {B.~G.}\ \bibnamefont {Sumpter}},
  \bibinfo {author} {\bibfnamefont {H.}~\bibnamefont {Terrones}}, \bibinfo
  {author} {\bibfnamefont {F.}~\bibnamefont {Xia}}, \bibinfo {author}
  {\bibfnamefont {Y.}~\bibnamefont {Wang}}, \bibinfo {author} {\bibfnamefont
  {J.}~\bibnamefont {Zhu}}, \bibinfo {author} {\bibfnamefont {D.}~\bibnamefont
  {Akinwande}}, \bibinfo {author} {\bibfnamefont {N.}~\bibnamefont {Alem}},
  \bibinfo {author} {\bibfnamefont {J.~A.}\ \bibnamefont {Schuller}}, \bibinfo
  {author} {\bibfnamefont {R.~E.}\ \bibnamefont {Schaak}}, \bibinfo {author}
  {\bibfnamefont {M.}~\bibnamefont {Terrones}}, \ and\ \bibinfo {author}
  {\bibfnamefont {J.~A.}\ \bibnamefont {Robinson}},\ }\href {\doibase
  10.1021/acsnano.5b05556} {\bibfield  {journal} {\bibinfo  {journal} {ACS
  Nano}\ }\textbf {\bibinfo {volume} {9}},\ \bibinfo {pages} {11509} (\bibinfo
  {year} {2015})}\BibitemShut {NoStop}%
\bibitem [{\citenamefont {Zibouche}\ \emph {et~al.}(2014)\citenamefont
  {Zibouche}, \citenamefont {Kuc}, \citenamefont {Musfeldt},\ and\
  \citenamefont {Heine}}]{Zibouche2014}%
  \BibitemOpen
  \bibfield  {author} {\bibinfo {author} {\bibfnamefont {N.}~\bibnamefont
  {Zibouche}}, \bibinfo {author} {\bibfnamefont {A.}~\bibnamefont {Kuc}},
  \bibinfo {author} {\bibfnamefont {J.}~\bibnamefont {Musfeldt}}, \ and\
  \bibinfo {author} {\bibfnamefont {T.}~\bibnamefont {Heine}},\ }\href
  {\doibase 10.1002/andp.201400137} {\bibfield  {journal} {\bibinfo  {journal}
  {Ann. Phys. (Berlin)}\ }\textbf {\bibinfo {volume} {526}},\ \bibinfo {pages}
  {395} (\bibinfo {year} {2014})}\BibitemShut {NoStop}%
\bibitem [{\citenamefont {Han}(2016)}]{Han2016}%
  \BibitemOpen
  \bibfield  {author} {\bibinfo {author} {\bibfnamefont {W.}~\bibnamefont
  {Han}},\ }\href {\doibase 10.1063/1.4941712} {\bibfield  {journal} {\bibinfo
  {journal} {APL Mater.}\ }\textbf {\bibinfo {volume} {4}},\ \bibinfo {pages}
  {032401} (\bibinfo {year} {2016})}\BibitemShut {NoStop}%
\bibitem [{\citenamefont {Xu}\ \emph {et~al.}(2014)\citenamefont {Xu},
  \citenamefont {Yao}, \citenamefont {Xiao},\ and\ \citenamefont
  {Heinz}}]{Xu2014NatPhys}%
  \BibitemOpen
  \bibfield  {author} {\bibinfo {author} {\bibfnamefont {X.}~\bibnamefont
  {Xu}}, \bibinfo {author} {\bibfnamefont {W.}~\bibnamefont {Yao}}, \bibinfo
  {author} {\bibfnamefont {D.}~\bibnamefont {Xiao}}, \ and\ \bibinfo {author}
  {\bibfnamefont {T.~F.}\ \bibnamefont {Heinz}},\ }\href {\doibase
  10.1038/nphys2942} {\bibfield  {journal} {\bibinfo  {journal} {Nat. Phys.}\
  }\textbf {\bibinfo {volume} {10}},\ \bibinfo {pages} {343} (\bibinfo {year}
  {2014})}\BibitemShut {NoStop}%
\bibitem [{\citenamefont {Liu}\ \emph {et~al.}(2015)\citenamefont {Liu},
  \citenamefont {Xiao}, \citenamefont {Yao}, \citenamefont {Xu},\ and\
  \citenamefont {Yao}}]{Liu2015}%
  \BibitemOpen
  \bibfield  {author} {\bibinfo {author} {\bibfnamefont {G.-B.}\ \bibnamefont
  {Liu}}, \bibinfo {author} {\bibfnamefont {D.}~\bibnamefont {Xiao}}, \bibinfo
  {author} {\bibfnamefont {Y.}~\bibnamefont {Yao}}, \bibinfo {author}
  {\bibfnamefont {X.}~\bibnamefont {Xu}}, \ and\ \bibinfo {author}
  {\bibfnamefont {W.}~\bibnamefont {Yao}},\ }\href {\doibase
  10.1039/C4CS00301B} {\bibfield  {journal} {\bibinfo  {journal} {Chem. Soc.
  Rev.}\ }\textbf {\bibinfo {volume} {44}},\ \bibinfo {pages} {2643} (\bibinfo
  {year} {2015})}\BibitemShut {NoStop}%
\bibitem [{\citenamefont {Castellanos-Gomez}(2016)}]{Castellanos2016}%
  \BibitemOpen
  \bibfield  {author} {\bibinfo {author} {\bibfnamefont {A.}~\bibnamefont
  {Castellanos-Gomez}},\ }\href {\doibase 10.1038/nphoton.2016.53} {\bibfield
  {journal} {\bibinfo  {journal} {Nat. Photonics}\ }\textbf {\bibinfo {volume}
  {10}},\ \bibinfo {pages} {202} (\bibinfo {year} {2016})}\BibitemShut
  {NoStop}%
\bibitem [{\citenamefont {Zhu}\ \emph {et~al.}(2011)\citenamefont {Zhu},
  \citenamefont {Cheng},\ and\ \citenamefont {Schwingenschl\"ogl}}]{Zhu2011}%
  \BibitemOpen
  \bibfield  {author} {\bibinfo {author} {\bibfnamefont {Z.~Y.}\ \bibnamefont
  {Zhu}}, \bibinfo {author} {\bibfnamefont {Y.~C.}\ \bibnamefont {Cheng}}, \
  and\ \bibinfo {author} {\bibfnamefont {U.}~\bibnamefont
  {Schwingenschl\"ogl}},\ }\href {\doibase 10.1103/PhysRevB.84.153402}
  {\bibfield  {journal} {\bibinfo  {journal} {Phys. Rev. B}\ }\textbf {\bibinfo
  {volume} {84}},\ \bibinfo {pages} {153402} (\bibinfo {year}
  {2011})}\BibitemShut {NoStop}%
\bibitem [{\citenamefont {Cheiwchanchamnangij}\ and\ \citenamefont
  {Lambrecht}(2012)}]{Cheiwchanchamnangij2012}%
  \BibitemOpen
  \bibfield  {author} {\bibinfo {author} {\bibfnamefont {T.}~\bibnamefont
  {Cheiwchanchamnangij}}\ and\ \bibinfo {author} {\bibfnamefont {W.~R.~L.}\
  \bibnamefont {Lambrecht}},\ }\href {\doibase 10.1103/PhysRevB.85.205302}
  {\bibfield  {journal} {\bibinfo  {journal} {Phys. Rev. B}\ }\textbf {\bibinfo
  {volume} {85}},\ \bibinfo {pages} {205302} (\bibinfo {year}
  {2012})}\BibitemShut {NoStop}%
\bibitem [{\citenamefont {Xiao}\ \emph {et~al.}(2012)\citenamefont {Xiao},
  \citenamefont {Liu}, \citenamefont {Feng}, \citenamefont {Xu},\ and\
  \citenamefont {Yao}}]{Xiao2012}%
  \BibitemOpen
  \bibfield  {author} {\bibinfo {author} {\bibfnamefont {D.}~\bibnamefont
  {Xiao}}, \bibinfo {author} {\bibfnamefont {G.-B.}\ \bibnamefont {Liu}},
  \bibinfo {author} {\bibfnamefont {W.}~\bibnamefont {Feng}}, \bibinfo {author}
  {\bibfnamefont {X.}~\bibnamefont {Xu}}, \ and\ \bibinfo {author}
  {\bibfnamefont {W.}~\bibnamefont {Yao}},\ }\href {\doibase
  10.1103/PhysRevLett.108.196802} {\bibfield  {journal} {\bibinfo  {journal}
  {Phys. Rev. Lett.}\ }\textbf {\bibinfo {volume} {108}},\ \bibinfo {pages}
  {196802} (\bibinfo {year} {2012})}\BibitemShut {NoStop}%
\bibitem [{\citenamefont {Mak}\ \emph {et~al.}(2010)\citenamefont {Mak},
  \citenamefont {Lee}, \citenamefont {Hone}, \citenamefont {Shan},\ and\
  \citenamefont {Heinz}}]{Mak2010}%
  \BibitemOpen
  \bibfield  {author} {\bibinfo {author} {\bibfnamefont {K.~F.}\ \bibnamefont
  {Mak}}, \bibinfo {author} {\bibfnamefont {C.}~\bibnamefont {Lee}}, \bibinfo
  {author} {\bibfnamefont {J.}~\bibnamefont {Hone}}, \bibinfo {author}
  {\bibfnamefont {J.}~\bibnamefont {Shan}}, \ and\ \bibinfo {author}
  {\bibfnamefont {T.~F.}\ \bibnamefont {Heinz}},\ }\href {\doibase
  10.1103/PhysRevLett.105.136805} {\bibfield  {journal} {\bibinfo  {journal}
  {Phys. Rev. Lett.}\ }\textbf {\bibinfo {volume} {105}},\ \bibinfo {pages}
  {136805} (\bibinfo {year} {2010})}\BibitemShut {NoStop}%
\bibitem [{\citenamefont {van~der Zande}\ \emph {et~al.}(2013)\citenamefont
  {van~der Zande}, \citenamefont {Huang}, \citenamefont {Chenet}, \citenamefont
  {Berkelbach}, \citenamefont {You}, \citenamefont {Lee}, \citenamefont
  {Heinz}, \citenamefont {Reichman}, \citenamefont {Muller},\ and\
  \citenamefont {Hone}}]{Van2013}%
  \BibitemOpen
  \bibfield  {author} {\bibinfo {author} {\bibfnamefont {A.~M.}\ \bibnamefont
  {van~der Zande}}, \bibinfo {author} {\bibfnamefont {P.~Y.}\ \bibnamefont
  {Huang}}, \bibinfo {author} {\bibfnamefont {D.~A.}\ \bibnamefont {Chenet}},
  \bibinfo {author} {\bibfnamefont {T.~C.}\ \bibnamefont {Berkelbach}},
  \bibinfo {author} {\bibfnamefont {Y.}~\bibnamefont {You}}, \bibinfo {author}
  {\bibfnamefont {G.-H.}\ \bibnamefont {Lee}}, \bibinfo {author} {\bibfnamefont
  {T.~F.}\ \bibnamefont {Heinz}}, \bibinfo {author} {\bibfnamefont {D.~R.}\
  \bibnamefont {Reichman}}, \bibinfo {author} {\bibfnamefont {D.~A.}\
  \bibnamefont {Muller}}, \ and\ \bibinfo {author} {\bibfnamefont {J.~C.}\
  \bibnamefont {Hone}},\ }\href {\doibase 10.1038/nmat3633} {\bibfield
  {journal} {\bibinfo  {journal} {Nat. Mater.}\ }\textbf {\bibinfo {volume}
  {12}},\ \bibinfo {pages} {554} (\bibinfo {year} {2013})}\BibitemShut
  {NoStop}%
\bibitem [{\citenamefont {Cao}\ \emph {et~al.}(2015)\citenamefont {Cao},
  \citenamefont {Shen}, \citenamefont {Liang}, \citenamefont {Chen},\ and\
  \citenamefont {Shu}}]{Cao2015hexagonalflakes}%
  \BibitemOpen
  \bibfield  {author} {\bibinfo {author} {\bibfnamefont {D.}~\bibnamefont
  {Cao}}, \bibinfo {author} {\bibfnamefont {T.}~\bibnamefont {Shen}}, \bibinfo
  {author} {\bibfnamefont {P.}~\bibnamefont {Liang}}, \bibinfo {author}
  {\bibfnamefont {X.}~\bibnamefont {Chen}}, \ and\ \bibinfo {author}
  {\bibfnamefont {H.}~\bibnamefont {Shu}},\ }\href {\doibase 10.1021/jp5097713}
  {\bibfield  {journal} {\bibinfo  {journal} {J. Phys. Chem. C}\ }\textbf
  {\bibinfo {volume} {119}},\ \bibinfo {pages} {4294} (\bibinfo {year}
  {2015})}\BibitemShut {NoStop}%
\bibitem [{\citenamefont {Wang}\ \emph {et~al.}(2013)\citenamefont {Wang},
  \citenamefont {Feng}, \citenamefont {Wu},\ and\ \citenamefont
  {Jiao}}]{Wang2013rhomboidflakes}%
  \BibitemOpen
  \bibfield  {author} {\bibinfo {author} {\bibfnamefont {X.}~\bibnamefont
  {Wang}}, \bibinfo {author} {\bibfnamefont {H.}~\bibnamefont {Feng}}, \bibinfo
  {author} {\bibfnamefont {Y.}~\bibnamefont {Wu}}, \ and\ \bibinfo {author}
  {\bibfnamefont {L.}~\bibnamefont {Jiao}},\ }\href {\doibase
  10.1021/ja4013485} {\bibfield  {journal} {\bibinfo  {journal} {J. Am. Chem.
  Soc.}\ }\textbf {\bibinfo {volume} {135}},\ \bibinfo {pages} {5304} (\bibinfo
  {year} {2013})}\BibitemShut {NoStop}%
\bibitem [{\citenamefont {Lauritsen}\ \emph {et~al.}(2007)\citenamefont
  {Lauritsen}, \citenamefont {Kibsgaard}, \citenamefont {Helveg}, \citenamefont
  {Tops{\o}e}, \citenamefont {Clausen}, \citenamefont {L{\ae}gsgaard},\ and\
  \citenamefont {Besenbacher}}]{Lauritsen2007}%
  \BibitemOpen
  \bibfield  {author} {\bibinfo {author} {\bibfnamefont {J.~V.}\ \bibnamefont
  {Lauritsen}}, \bibinfo {author} {\bibfnamefont {J.}~\bibnamefont
  {Kibsgaard}}, \bibinfo {author} {\bibfnamefont {S.}~\bibnamefont {Helveg}},
  \bibinfo {author} {\bibfnamefont {H.}~\bibnamefont {Tops{\o}e}}, \bibinfo
  {author} {\bibfnamefont {B.~S.}\ \bibnamefont {Clausen}}, \bibinfo {author}
  {\bibfnamefont {E.}~\bibnamefont {L{\ae}gsgaard}}, \ and\ \bibinfo {author}
  {\bibfnamefont {F.}~\bibnamefont {Besenbacher}},\ }\href {\doibase
  10.1038/nnano.2006.171} {\bibfield  {journal} {\bibinfo  {journal} {Nat.
  Nanotechnol.}\ }\textbf {\bibinfo {volume} {2}},\ \bibinfo {pages} {53}
  (\bibinfo {year} {2007})}\BibitemShut {NoStop}%
\bibitem [{\citenamefont {Chiu}\ \emph {et~al.}(2014)\citenamefont {Chiu},
  \citenamefont {Li}, \citenamefont {Zhang}, \citenamefont {Hsu}, \citenamefont
  {Chang}, \citenamefont {Terrones}, \citenamefont {Terrones},\ and\
  \citenamefont {Li}}]{Chiu2014}%
  \BibitemOpen
  \bibfield  {author} {\bibinfo {author} {\bibfnamefont {M.-H.}\ \bibnamefont
  {Chiu}}, \bibinfo {author} {\bibfnamefont {M.-Y.}\ \bibnamefont {Li}},
  \bibinfo {author} {\bibfnamefont {W.}~\bibnamefont {Zhang}}, \bibinfo
  {author} {\bibfnamefont {W.-T.}\ \bibnamefont {Hsu}}, \bibinfo {author}
  {\bibfnamefont {W.-H.}\ \bibnamefont {Chang}}, \bibinfo {author}
  {\bibfnamefont {M.}~\bibnamefont {Terrones}}, \bibinfo {author}
  {\bibfnamefont {H.}~\bibnamefont {Terrones}}, \ and\ \bibinfo {author}
  {\bibfnamefont {L.-J.}\ \bibnamefont {Li}},\ }\href {\doibase
  10.1021/nn504229z} {\bibfield  {journal} {\bibinfo  {journal} {ACS Nano}\
  }\textbf {\bibinfo {volume} {8}},\ \bibinfo {pages} {9649} (\bibinfo {year}
  {2014})}\BibitemShut {NoStop}%
\bibitem [{\citenamefont {Li}\ \emph {et~al.}(2008)\citenamefont {Li},
  \citenamefont {Zhou}, \citenamefont {Zhang},\ and\ \citenamefont
  {Chen}}]{Li2008}%
  \BibitemOpen
  \bibfield  {author} {\bibinfo {author} {\bibfnamefont {Y.}~\bibnamefont
  {Li}}, \bibinfo {author} {\bibfnamefont {Z.}~\bibnamefont {Zhou}}, \bibinfo
  {author} {\bibfnamefont {S.}~\bibnamefont {Zhang}}, \ and\ \bibinfo {author}
  {\bibfnamefont {Z.}~\bibnamefont {Chen}},\ }\href {\doibase
  10.1021/ja805545x} {\bibfield  {journal} {\bibinfo  {journal} {J. Am. Chem.
  Soc.}\ }\textbf {\bibinfo {volume} {130}},\ \bibinfo {pages} {16739}
  (\bibinfo {year} {2008})}\BibitemShut {NoStop}%
\bibitem [{\citenamefont {Botello-M\'endez}\ \emph {et~al.}(2009)\citenamefont
  {Botello-M\'endez}, \citenamefont {L\'opez-Ur\'ias}, \citenamefont
  {Terrones},\ and\ \citenamefont {Terrones}}]{Botello2009}%
  \BibitemOpen
  \bibfield  {author} {\bibinfo {author} {\bibfnamefont {A.~R.}\ \bibnamefont
  {Botello-M\'endez}}, \bibinfo {author} {\bibfnamefont {F.}~\bibnamefont
  {L\'opez-Ur\'ias}}, \bibinfo {author} {\bibfnamefont {M.}~\bibnamefont
  {Terrones}}, \ and\ \bibinfo {author} {\bibfnamefont {H.}~\bibnamefont
  {Terrones}},\ }\href {\doibase 10.1088/0957-4484/20/32/325703} {\bibfield
  {journal} {\bibinfo  {journal} {Nanotechnology}\ }\textbf {\bibinfo {volume}
  {20}},\ \bibinfo {pages} {325703} (\bibinfo {year} {2009})}\BibitemShut
  {NoStop}%
\bibitem [{\citenamefont {Tongay}\ \emph {et~al.}(2012)\citenamefont {Tongay},
  \citenamefont {Varnoosfaderani}, \citenamefont {Appleton}, \citenamefont
  {Wu},\ and\ \citenamefont {Hebard}}]{Tongay2012}%
  \BibitemOpen
  \bibfield  {author} {\bibinfo {author} {\bibfnamefont {S.}~\bibnamefont
  {Tongay}}, \bibinfo {author} {\bibfnamefont {S.~S.}\ \bibnamefont
  {Varnoosfaderani}}, \bibinfo {author} {\bibfnamefont {B.~R.}\ \bibnamefont
  {Appleton}}, \bibinfo {author} {\bibfnamefont {J.}~\bibnamefont {Wu}}, \ and\
  \bibinfo {author} {\bibfnamefont {A.~F.}\ \bibnamefont {Hebard}},\ }\href
  {\doibase 10.1063/1.4753797} {\bibfield  {journal} {\bibinfo  {journal}
  {Appl. Phys. Lett.}\ }\textbf {\bibinfo {volume} {101}},\ \bibinfo {pages}
  {123105} (\bibinfo {year} {2012})}\BibitemShut {NoStop}%
\bibitem [{\citenamefont {Ruderman}\ and\ \citenamefont
  {Kittel}(1954)}]{RudermanKittel1954}%
  \BibitemOpen
  \bibfield  {author} {\bibinfo {author} {\bibfnamefont {M.~A.}\ \bibnamefont
  {Ruderman}}\ and\ \bibinfo {author} {\bibfnamefont {C.}~\bibnamefont
  {Kittel}},\ }\href {\doibase 10.1103/PhysRev.96.99} {\bibfield  {journal}
  {\bibinfo  {journal} {Phys. Rev.}\ }\textbf {\bibinfo {volume} {96}},\
  \bibinfo {pages} {99} (\bibinfo {year} {1954})}\BibitemShut {NoStop}%
\bibitem [{\citenamefont {Kasuya}(1956)}]{Kasuya1956}%
  \BibitemOpen
  \bibfield  {author} {\bibinfo {author} {\bibfnamefont {T.}~\bibnamefont
  {Kasuya}},\ }\href {\doibase 10.1143/PTP.16.45} {\bibfield  {journal}
  {\bibinfo  {journal} {Progr. Theor. Phys.}\ }\textbf {\bibinfo {volume}
  {16}},\ \bibinfo {pages} {45} (\bibinfo {year} {1956})}\BibitemShut {NoStop}%
\bibitem [{\citenamefont {Yosida}(1957)}]{Yosida1957}%
  \BibitemOpen
  \bibfield  {author} {\bibinfo {author} {\bibfnamefont {K.}~\bibnamefont
  {Yosida}},\ }\href {\doibase 10.1103/PhysRev.106.893} {\bibfield  {journal}
  {\bibinfo  {journal} {Phys. Rev.}\ }\textbf {\bibinfo {volume} {106}},\
  \bibinfo {pages} {893} (\bibinfo {year} {1957})}\BibitemShut {NoStop}%
\bibitem [{\citenamefont {Laskar}\ \emph {et~al.}(2014)\citenamefont {Laskar},
  \citenamefont {Nath}, \citenamefont {Ma}, \citenamefont {Lee}, \citenamefont
  {Lee}, \citenamefont {Kent}, \citenamefont {Yang}, \citenamefont {Mishra},
  \citenamefont {Roldan}, \citenamefont {Idrobo}, \citenamefont {Pantelides},
  \citenamefont {Pennycook}, \citenamefont {Myers}, \citenamefont {Wu},\ and\
  \citenamefont {Rajan}}]{Laskar2014}%
  \BibitemOpen
  \bibfield  {author} {\bibinfo {author} {\bibfnamefont {M.~R.}\ \bibnamefont
  {Laskar}}, \bibinfo {author} {\bibfnamefont {D.~N.}\ \bibnamefont {Nath}},
  \bibinfo {author} {\bibfnamefont {L.}~\bibnamefont {Ma}}, \bibinfo {author}
  {\bibfnamefont {E.~W.}\ \bibnamefont {Lee}}, \bibinfo {author} {\bibfnamefont
  {C.~H.}\ \bibnamefont {Lee}}, \bibinfo {author} {\bibfnamefont
  {T.}~\bibnamefont {Kent}}, \bibinfo {author} {\bibfnamefont {Z.}~\bibnamefont
  {Yang}}, \bibinfo {author} {\bibfnamefont {R.}~\bibnamefont {Mishra}},
  \bibinfo {author} {\bibfnamefont {M.~A.}\ \bibnamefont {Roldan}}, \bibinfo
  {author} {\bibfnamefont {J.-C.}\ \bibnamefont {Idrobo}}, \bibinfo {author}
  {\bibfnamefont {S.~T.}\ \bibnamefont {Pantelides}}, \bibinfo {author}
  {\bibfnamefont {S.~J.}\ \bibnamefont {Pennycook}}, \bibinfo {author}
  {\bibfnamefont {R.~C.}\ \bibnamefont {Myers}}, \bibinfo {author}
  {\bibfnamefont {Y.}~\bibnamefont {Wu}}, \ and\ \bibinfo {author}
  {\bibfnamefont {S.}~\bibnamefont {Rajan}},\ }\href {\doibase
  10.1063/1.4867197} {\bibfield  {journal} {\bibinfo  {journal} {Appl. Phys.
  Lett.}\ }\textbf {\bibinfo {volume} {104}},\ \bibinfo {pages} {092104}
  (\bibinfo {year} {2014})}\BibitemShut {NoStop}%
\bibitem [{\citenamefont {Suh}\ \emph {et~al.}(2014)\citenamefont {Suh},
  \citenamefont {Park}, \citenamefont {Lin}, \citenamefont {Fu}, \citenamefont
  {Park}, \citenamefont {Jung}, \citenamefont {Chen}, \citenamefont {Ko},
  \citenamefont {Jang}, \citenamefont {Sun}, \citenamefont {Sinclair},
  \citenamefont {Chang}, \citenamefont {Tongay},\ and\ \citenamefont
  {Wu}}]{Suh2014}%
  \BibitemOpen
  \bibfield  {author} {\bibinfo {author} {\bibfnamefont {J.}~\bibnamefont
  {Suh}}, \bibinfo {author} {\bibfnamefont {T.-E.}\ \bibnamefont {Park}},
  \bibinfo {author} {\bibfnamefont {D.-Y.}\ \bibnamefont {Lin}}, \bibinfo
  {author} {\bibfnamefont {D.}~\bibnamefont {Fu}}, \bibinfo {author}
  {\bibfnamefont {J.}~\bibnamefont {Park}}, \bibinfo {author} {\bibfnamefont
  {H.~J.}\ \bibnamefont {Jung}}, \bibinfo {author} {\bibfnamefont
  {Y.}~\bibnamefont {Chen}}, \bibinfo {author} {\bibfnamefont {C.}~\bibnamefont
  {Ko}}, \bibinfo {author} {\bibfnamefont {C.}~\bibnamefont {Jang}}, \bibinfo
  {author} {\bibfnamefont {Y.}~\bibnamefont {Sun}}, \bibinfo {author}
  {\bibfnamefont {R.}~\bibnamefont {Sinclair}}, \bibinfo {author}
  {\bibfnamefont {J.}~\bibnamefont {Chang}}, \bibinfo {author} {\bibfnamefont
  {S.}~\bibnamefont {Tongay}}, \ and\ \bibinfo {author} {\bibfnamefont
  {J.}~\bibnamefont {Wu}},\ }\href {\doibase 10.1021/nl503251h} {\bibfield
  {journal} {\bibinfo  {journal} {Nano Letters}\ }\textbf {\bibinfo {volume}
  {14}},\ \bibinfo {pages} {6976} (\bibinfo {year} {2014})}\BibitemShut
  {NoStop}%
\bibitem [{\citenamefont {Nipane}\ \emph {et~al.}(2016)\citenamefont {Nipane},
  \citenamefont {Karmakar}, \citenamefont {Kaushik}, \citenamefont {Karande},\
  and\ \citenamefont {Lodha}}]{Nipane2016}%
  \BibitemOpen
  \bibfield  {author} {\bibinfo {author} {\bibfnamefont {A.}~\bibnamefont
  {Nipane}}, \bibinfo {author} {\bibfnamefont {D.}~\bibnamefont {Karmakar}},
  \bibinfo {author} {\bibfnamefont {N.}~\bibnamefont {Kaushik}}, \bibinfo
  {author} {\bibfnamefont {S.}~\bibnamefont {Karande}}, \ and\ \bibinfo
  {author} {\bibfnamefont {S.}~\bibnamefont {Lodha}},\ }\href {\doibase
  10.1021/acsnano.5b06529} {\bibfield  {journal} {\bibinfo  {journal} {ACS
  Nano}\ }\textbf {\bibinfo {volume} {10}},\ \bibinfo {pages} {2128} (\bibinfo
  {year} {2016})}\BibitemShut {NoStop}%
\bibitem [{\citenamefont {Dolui}\ \emph {et~al.}(2013)\citenamefont {Dolui},
  \citenamefont {Rungger}, \citenamefont {Das~Pemmaraju},\ and\ \citenamefont
  {Sanvito}}]{Dolui2013}%
  \BibitemOpen
  \bibfield  {author} {\bibinfo {author} {\bibfnamefont {K.}~\bibnamefont
  {Dolui}}, \bibinfo {author} {\bibfnamefont {I.}~\bibnamefont {Rungger}},
  \bibinfo {author} {\bibfnamefont {C.}~\bibnamefont {Das~Pemmaraju}}, \ and\
  \bibinfo {author} {\bibfnamefont {S.}~\bibnamefont {Sanvito}},\ }\href
  {\doibase 10.1103/PhysRevB.88.075420} {\bibfield  {journal} {\bibinfo
  {journal} {Phys. Rev. B}\ }\textbf {\bibinfo {volume} {88}},\ \bibinfo
  {pages} {075420} (\bibinfo {year} {2013})}\BibitemShut {NoStop}%
\bibitem [{\citenamefont {Mishra}\ \emph {et~al.}(2013)\citenamefont {Mishra},
  \citenamefont {Zhou}, \citenamefont {Pennycook}, \citenamefont {Pantelides},\
  and\ \citenamefont {Idrobo}}]{Mishra2013}%
  \BibitemOpen
  \bibfield  {author} {\bibinfo {author} {\bibfnamefont {R.}~\bibnamefont
  {Mishra}}, \bibinfo {author} {\bibfnamefont {W.}~\bibnamefont {Zhou}},
  \bibinfo {author} {\bibfnamefont {S.~J.}\ \bibnamefont {Pennycook}}, \bibinfo
  {author} {\bibfnamefont {S.~T.}\ \bibnamefont {Pantelides}}, \ and\ \bibinfo
  {author} {\bibfnamefont {J.-C.}\ \bibnamefont {Idrobo}},\ }\href {\doibase
  10.1103/PhysRevB.88.144409} {\bibfield  {journal} {\bibinfo  {journal} {Phys.
  Rev. B}\ }\textbf {\bibinfo {volume} {88}},\ \bibinfo {pages} {144409}
  (\bibinfo {year} {2013})}\BibitemShut {NoStop}%
\bibitem [{\citenamefont {Cheng}\ \emph {et~al.}(2013)\citenamefont {Cheng},
  \citenamefont {Zhu}, \citenamefont {Mi}, \citenamefont {Guo},\ and\
  \citenamefont {Schwingenschl\"ogl}}]{Cheng2013}%
  \BibitemOpen
  \bibfield  {author} {\bibinfo {author} {\bibfnamefont {Y.~C.}\ \bibnamefont
  {Cheng}}, \bibinfo {author} {\bibfnamefont {Z.~Y.}\ \bibnamefont {Zhu}},
  \bibinfo {author} {\bibfnamefont {W.~B.}\ \bibnamefont {Mi}}, \bibinfo
  {author} {\bibfnamefont {Z.~B.}\ \bibnamefont {Guo}}, \ and\ \bibinfo
  {author} {\bibfnamefont {U.}~\bibnamefont {Schwingenschl\"ogl}},\ }\href
  {\doibase 10.1103/PhysRevB.87.100401} {\bibfield  {journal} {\bibinfo
  {journal} {Phys. Rev. B}\ }\textbf {\bibinfo {volume} {87}},\ \bibinfo
  {pages} {100401} (\bibinfo {year} {2013})}\BibitemShut {NoStop}%
\bibitem [{\citenamefont {Khajetoorians}\ \emph {et~al.}(2012)\citenamefont
  {Khajetoorians}, \citenamefont {Wiebe}, \citenamefont {Chilian},
  \citenamefont {Lounis}, \citenamefont {Bl{\"u}gel},\ and\ \citenamefont
  {Wiesendanger}}]{Khajetoorians2012}%
  \BibitemOpen
  \bibfield  {author} {\bibinfo {author} {\bibfnamefont {A.~A.}\ \bibnamefont
  {Khajetoorians}}, \bibinfo {author} {\bibfnamefont {J.}~\bibnamefont
  {Wiebe}}, \bibinfo {author} {\bibfnamefont {B.}~\bibnamefont {Chilian}},
  \bibinfo {author} {\bibfnamefont {S.}~\bibnamefont {Lounis}}, \bibinfo
  {author} {\bibfnamefont {S.}~\bibnamefont {Bl{\"u}gel}}, \ and\ \bibinfo
  {author} {\bibfnamefont {R.}~\bibnamefont {Wiesendanger}},\ }\href {\doibase
  10.1038/nphys2299} {\bibfield  {journal} {\bibinfo  {journal} {Nat. Phys.}\
  }\textbf {\bibinfo {volume} {8}},\ \bibinfo {pages} {497} (\bibinfo {year}
  {2012})}\BibitemShut {NoStop}%
\bibitem [{\citenamefont {Lounis}(2014)}]{Lounis2014}%
  \BibitemOpen
  \bibfield  {author} {\bibinfo {author} {\bibfnamefont {S.}~\bibnamefont
  {Lounis}},\ }\href {\doibase 10.1088/0953-8984/26/27/273201} {\bibfield
  {journal} {\bibinfo  {journal} {J. Phys.: Condens. Matter}\ }\textbf
  {\bibinfo {volume} {26}},\ \bibinfo {pages} {273201} (\bibinfo {year}
  {2014})}\BibitemShut {NoStop}%
\bibitem [{\citenamefont {Cong}\ \emph {et~al.}(2015)\citenamefont {Cong},
  \citenamefont {Tang}, \citenamefont {Zhao},\ and\ \citenamefont
  {Chu}}]{Cong2015}%
  \BibitemOpen
  \bibfield  {author} {\bibinfo {author} {\bibfnamefont {W.~T.}\ \bibnamefont
  {Cong}}, \bibinfo {author} {\bibfnamefont {Z.}~\bibnamefont {Tang}}, \bibinfo
  {author} {\bibfnamefont {X.~G.}\ \bibnamefont {Zhao}}, \ and\ \bibinfo
  {author} {\bibfnamefont {J.~H.}\ \bibnamefont {Chu}},\ }\href {\doibase
  10.1038/srep09361} {\bibfield  {journal} {\bibinfo  {journal} {Sci. Rep.}\
  }\textbf {\bibinfo {volume} {5}},\ \bibinfo {pages} {9361} (\bibinfo {year}
  {2015})}\BibitemShut {NoStop}%
\bibitem [{\citenamefont {Lu}\ and\ \citenamefont {Leburton}(2014)}]{Lu2014}%
  \BibitemOpen
  \bibfield  {author} {\bibinfo {author} {\bibfnamefont {S.-C.}\ \bibnamefont
  {Lu}}\ and\ \bibinfo {author} {\bibfnamefont {J.-P.}\ \bibnamefont
  {Leburton}},\ }\href {\doibase 10.1186/1556-276X-9-676} {\bibfield  {journal}
  {\bibinfo  {journal} {Nanoscale Res. Lett.}\ }\textbf {\bibinfo {volume}
  {9}},\ \bibinfo {pages} {676} (\bibinfo {year} {2014})}\BibitemShut {NoStop}%
\bibitem [{\citenamefont {Saab}\ and\ \citenamefont
  {Raybaud}(2016)}]{Saab2016}%
  \BibitemOpen
  \bibfield  {author} {\bibinfo {author} {\bibfnamefont {M.}~\bibnamefont
  {Saab}}\ and\ \bibinfo {author} {\bibfnamefont {P.}~\bibnamefont {Raybaud}},\
  }\href {\doibase 10.1021/acs.jpcc.6b02865} {\bibfield  {journal} {\bibinfo
  {journal} {J. Phys. Chem. C}\ }\textbf {\bibinfo {volume} {120}},\ \bibinfo
  {pages} {10691} (\bibinfo {year} {2016})}\BibitemShut {NoStop}%
\bibitem [{\citenamefont {Zhang}\ \emph {et~al.}(2015)\citenamefont {Zhang},
  \citenamefont {Feng}, \citenamefont {Wang}, \citenamefont {Azcatl},
  \citenamefont {Lu}, \citenamefont {Addou}, \citenamefont {Wang},
  \citenamefont {Zhou}, \citenamefont {Lerach}, \citenamefont {Bojan},
  \citenamefont {Kim}, \citenamefont {Chen}, \citenamefont {Wallace},
  \citenamefont {Terrones}, \citenamefont {Zhu},\ and\ \citenamefont
  {Robinson}}]{Zhang2015}%
  \BibitemOpen
  \bibfield  {author} {\bibinfo {author} {\bibfnamefont {K.}~\bibnamefont
  {Zhang}}, \bibinfo {author} {\bibfnamefont {S.}~\bibnamefont {Feng}},
  \bibinfo {author} {\bibfnamefont {J.}~\bibnamefont {Wang}}, \bibinfo {author}
  {\bibfnamefont {A.}~\bibnamefont {Azcatl}}, \bibinfo {author} {\bibfnamefont
  {N.}~\bibnamefont {Lu}}, \bibinfo {author} {\bibfnamefont {R.}~\bibnamefont
  {Addou}}, \bibinfo {author} {\bibfnamefont {N.}~\bibnamefont {Wang}},
  \bibinfo {author} {\bibfnamefont {C.}~\bibnamefont {Zhou}}, \bibinfo {author}
  {\bibfnamefont {J.}~\bibnamefont {Lerach}}, \bibinfo {author} {\bibfnamefont
  {V.}~\bibnamefont {Bojan}}, \bibinfo {author} {\bibfnamefont {M.~J.}\
  \bibnamefont {Kim}}, \bibinfo {author} {\bibfnamefont {L.-Q.}\ \bibnamefont
  {Chen}}, \bibinfo {author} {\bibfnamefont {R.~M.}\ \bibnamefont {Wallace}},
  \bibinfo {author} {\bibfnamefont {M.}~\bibnamefont {Terrones}}, \bibinfo
  {author} {\bibfnamefont {J.}~\bibnamefont {Zhu}}, \ and\ \bibinfo {author}
  {\bibfnamefont {J.~A.}\ \bibnamefont {Robinson}},\ }\href {\doibase
  10.1021/acs.nanolett.5b02315} {\bibfield  {journal} {\bibinfo  {journal}
  {Nano Letters}\ }\textbf {\bibinfo {volume} {15}},\ \bibinfo {pages} {6586}
  (\bibinfo {year} {2015})}\BibitemShut {NoStop}%
\bibitem [{\citenamefont {Wang}\ \emph {et~al.}(2016)\citenamefont {Wang},
  \citenamefont {Sun}, \citenamefont {Yang}, \citenamefont {Li}, \citenamefont
  {Zhao}, \citenamefont {Xu}, \citenamefont {Zhang},\ and\ \citenamefont
  {Zeng}}]{Wang2016}%
  \BibitemOpen
  \bibfield  {author} {\bibinfo {author} {\bibfnamefont {J.}~\bibnamefont
  {Wang}}, \bibinfo {author} {\bibfnamefont {F.}~\bibnamefont {Sun}}, \bibinfo
  {author} {\bibfnamefont {S.}~\bibnamefont {Yang}}, \bibinfo {author}
  {\bibfnamefont {Y.}~\bibnamefont {Li}}, \bibinfo {author} {\bibfnamefont
  {C.}~\bibnamefont {Zhao}}, \bibinfo {author} {\bibfnamefont {M.}~\bibnamefont
  {Xu}}, \bibinfo {author} {\bibfnamefont {Y.}~\bibnamefont {Zhang}}, \ and\
  \bibinfo {author} {\bibfnamefont {H.}~\bibnamefont {Zeng}},\ }\href {\doibase
  10.1063/1.4961883} {\bibfield  {journal} {\bibinfo  {journal} {Appl. Phys.
  Lett.}\ }\textbf {\bibinfo {volume} {109}},\ \bibinfo {pages} {092401}
  (\bibinfo {year} {2016})}\BibitemShut {NoStop}%
\bibitem [{\citenamefont {Parhizgar}\ \emph {et~al.}(2013)\citenamefont
  {Parhizgar}, \citenamefont {Rostami},\ and\ \citenamefont
  {Asgari}}]{Parhizgar2013}%
  \BibitemOpen
  \bibfield  {author} {\bibinfo {author} {\bibfnamefont {F.}~\bibnamefont
  {Parhizgar}}, \bibinfo {author} {\bibfnamefont {H.}~\bibnamefont {Rostami}},
  \ and\ \bibinfo {author} {\bibfnamefont {R.}~\bibnamefont {Asgari}},\ }\href
  {\doibase 10.1103/PhysRevB.87.125401} {\bibfield  {journal} {\bibinfo
  {journal} {Phys. Rev. B}\ }\textbf {\bibinfo {volume} {87}},\ \bibinfo
  {pages} {125401} (\bibinfo {year} {2013})}\BibitemShut {NoStop}%
\bibitem [{\citenamefont {Hatami}\ \emph {et~al.}(2014)\citenamefont {Hatami},
  \citenamefont {Kernreiter},\ and\ \citenamefont {Z\"ulicke}}]{Hatami2014}%
  \BibitemOpen
  \bibfield  {author} {\bibinfo {author} {\bibfnamefont {H.}~\bibnamefont
  {Hatami}}, \bibinfo {author} {\bibfnamefont {T.}~\bibnamefont {Kernreiter}},
  \ and\ \bibinfo {author} {\bibfnamefont {U.}~\bibnamefont {Z\"ulicke}},\
  }\href {\doibase 10.1103/PhysRevB.90.045412} {\bibfield  {journal} {\bibinfo
  {journal} {Phys. Rev. B}\ }\textbf {\bibinfo {volume} {90}},\ \bibinfo
  {pages} {045412} (\bibinfo {year} {2014})}\BibitemShut {NoStop}%
\bibitem [{\citenamefont {Mastrogiuseppe}\ \emph {et~al.}(2014)\citenamefont
  {Mastrogiuseppe}, \citenamefont {Sandler},\ and\ \citenamefont
  {Ulloa}}]{Mastrogiuseppe2014}%
  \BibitemOpen
  \bibfield  {author} {\bibinfo {author} {\bibfnamefont {D.}~\bibnamefont
  {Mastrogiuseppe}}, \bibinfo {author} {\bibfnamefont {N.}~\bibnamefont
  {Sandler}}, \ and\ \bibinfo {author} {\bibfnamefont {S.~E.}\ \bibnamefont
  {Ulloa}},\ }\href {\doibase 10.1103/PhysRevB.90.161403} {\bibfield  {journal}
  {\bibinfo  {journal} {Phys. Rev. B}\ }\textbf {\bibinfo {volume} {90}},\
  \bibinfo {pages} {161403} (\bibinfo {year} {2014})}\BibitemShut {NoStop}%
\bibitem [{\citenamefont {\'Avalos-Ovando}\ \emph
  {et~al.}(2016{\natexlab{a}})\citenamefont {\'Avalos-Ovando}, \citenamefont
  {Mastrogiuseppe},\ and\ \citenamefont {Ulloa}}]{Avalos2016}%
  \BibitemOpen
  \bibfield  {author} {\bibinfo {author} {\bibfnamefont {O.}~\bibnamefont
  {\'Avalos-Ovando}}, \bibinfo {author} {\bibfnamefont {D.}~\bibnamefont
  {Mastrogiuseppe}}, \ and\ \bibinfo {author} {\bibfnamefont {S.~E.}\
  \bibnamefont {Ulloa}},\ }\href {\doibase 10.1103/PhysRevB.93.161404}
  {\bibfield  {journal} {\bibinfo  {journal} {Phys. Rev. B}\ }\textbf {\bibinfo
  {volume} {93}},\ \bibinfo {pages} {161404} (\bibinfo {year}
  {2016}{\natexlab{a}})}\BibitemShut {NoStop}%
\bibitem [{\citenamefont {\'Avalos-Ovando}\ \emph
  {et~al.}(2016{\natexlab{b}})\citenamefont {\'Avalos-Ovando}, \citenamefont
  {Mastrogiuseppe},\ and\ \citenamefont {Ulloa}}]{Avalos2016arxiv}%
  \BibitemOpen
  \bibfield  {author} {\bibinfo {author} {\bibfnamefont {O.}~\bibnamefont
  {\'Avalos-Ovando}}, \bibinfo {author} {\bibfnamefont {D.}~\bibnamefont
  {Mastrogiuseppe}}, \ and\ \bibinfo {author} {\bibfnamefont {S.~E.}\
  \bibnamefont {Ulloa}},\ }\href@noop {} {} (\bibinfo {year}
  {2016}{\natexlab{b}}),\ \Eprint {http://arxiv.org/abs/1607.08553}
  {arXiv:1607.08553} \BibitemShut {NoStop}%
\bibitem [{\citenamefont {Mi}\ \emph {et~al.}(2011)\citenamefont {Mi},
  \citenamefont {Yuan},\ and\ \citenamefont {Lyu}}]{Mi2011}%
  \BibitemOpen
  \bibfield  {author} {\bibinfo {author} {\bibfnamefont {S.}~\bibnamefont
  {Mi}}, \bibinfo {author} {\bibfnamefont {S.-H.}\ \bibnamefont {Yuan}}, \ and\
  \bibinfo {author} {\bibfnamefont {P.}~\bibnamefont {Lyu}},\ }\href {\doibase
  http://dx.doi.org/10.1063/1.3575338} {\bibfield  {journal} {\bibinfo
  {journal} {J. Appl. Phys.}\ }\textbf {\bibinfo {volume} {109}},\ \bibinfo
  {pages} {083931} (\bibinfo {year} {2011})}\BibitemShut {NoStop}%
\bibitem [{\citenamefont {Power}\ and\ \citenamefont
  {Ferreira}(2013)}]{Power2013}%
  \BibitemOpen
  \bibfield  {author} {\bibinfo {author} {\bibfnamefont {S.~R.}\ \bibnamefont
  {Power}}\ and\ \bibinfo {author} {\bibfnamefont {M.~S.}\ \bibnamefont
  {Ferreira}},\ }\href {\doibase 10.3390/cryst3010049} {\bibfield  {journal}
  {\bibinfo  {journal} {Crystals}\ }\textbf {\bibinfo {volume} {3}},\ \bibinfo
  {pages} {49} (\bibinfo {year} {2013})}\BibitemShut {NoStop}%
\bibitem [{\citenamefont {Kogan}(2011)}]{Kogan2011}%
  \BibitemOpen
  \bibfield  {author} {\bibinfo {author} {\bibfnamefont {E.}~\bibnamefont
  {Kogan}},\ }\href {\doibase 10.1103/PhysRevB.84.115119} {\bibfield  {journal}
  {\bibinfo  {journal} {Phys. Rev. B}\ }\textbf {\bibinfo {volume} {84}},\
  \bibinfo {pages} {115119} (\bibinfo {year} {2011})}\BibitemShut {NoStop}%
\bibitem [{\citenamefont {Sza\l{}owski}(2011)}]{Szalowski2011}%
  \BibitemOpen
  \bibfield  {author} {\bibinfo {author} {\bibfnamefont {K.}~\bibnamefont
  {Sza\l{}owski}},\ }\href {\doibase 10.1103/PhysRevB.84.205409} {\bibfield
  {journal} {\bibinfo  {journal} {Phys. Rev. B}\ }\textbf {\bibinfo {volume}
  {84}},\ \bibinfo {pages} {205409} (\bibinfo {year} {2011})}\BibitemShut
  {NoStop}%
\bibitem [{\citenamefont {Sza\l{}owski}(2013{\natexlab{a}})}]{Szalowski2013}%
  \BibitemOpen
  \bibfield  {author} {\bibinfo {author} {\bibfnamefont {K.}~\bibnamefont
  {Sza\l{}owski}},\ }\href {\doibase 10.1016/j.physe.2013.03.017} {\bibfield
  {journal} {\bibinfo  {journal} {Physica E}\ }\textbf {\bibinfo {volume}
  {52}},\ \bibinfo {pages} {46} (\bibinfo {year}
  {2013}{\natexlab{a}})}\BibitemShut {NoStop}%
\bibitem [{\citenamefont {Nikoofard}\ and\ \citenamefont
  {Semiromi}(2016)}]{Nikoofard2016}%
  \BibitemOpen
  \bibfield  {author} {\bibinfo {author} {\bibfnamefont {H.}~\bibnamefont
  {Nikoofard}}\ and\ \bibinfo {author} {\bibfnamefont {E.~H.}\ \bibnamefont
  {Semiromi}},\ }\href {\doibase 10.1140/epjb/e2016-70094-6} {\bibfield
  {journal} {\bibinfo  {journal} {Eur. Phys. J. B}\ }\textbf {\bibinfo {volume}
  {89}},\ \bibinfo {pages} {221} (\bibinfo {year} {2016})}\BibitemShut
  {NoStop}%
\bibitem [{\citenamefont
  {Sza\l{}owski}(2013{\natexlab{b}})}]{Szalowski2013jofcm}%
  \BibitemOpen
  \bibfield  {author} {\bibinfo {author} {\bibfnamefont {K.}~\bibnamefont
  {Sza\l{}owski}},\ }\href {\doibase 10.1088/0953-8984/25/16/166001} {\bibfield
   {journal} {\bibinfo  {journal} {J. Phys.: Condens. Matter}\ }\textbf
  {\bibinfo {volume} {25}},\ \bibinfo {pages} {166001} (\bibinfo {year}
  {2013}{\natexlab{b}})}\BibitemShut {NoStop}%
\bibitem [{\citenamefont {Black-Schaffer}(2010)}]{Black2010paper1}%
  \BibitemOpen
  \bibfield  {author} {\bibinfo {author} {\bibfnamefont {A.~M.}\ \bibnamefont
  {Black-Schaffer}},\ }\href {\doibase 10.1103/PhysRevB.81.205416} {\bibfield
  {journal} {\bibinfo  {journal} {Phys. Rev. B}\ }\textbf {\bibinfo {volume}
  {81}},\ \bibinfo {pages} {205416} (\bibinfo {year} {2010})}\BibitemShut
  {NoStop}%
\bibitem [{\citenamefont {Akbari-Sharbaf}\ and\ \citenamefont
  {Cottam}(2014)}]{Akbari2014}%
  \BibitemOpen
  \bibfield  {author} {\bibinfo {author} {\bibfnamefont {A.}~\bibnamefont
  {Akbari-Sharbaf}}\ and\ \bibinfo {author} {\bibfnamefont {M.~G.}\
  \bibnamefont {Cottam}},\ }\href {\doibase 10.1063/1.4902146} {\bibfield
  {journal} {\bibinfo  {journal} {J. Appl. Phys.}\ }\textbf {\bibinfo {volume}
  {116}},\ \bibinfo {pages} {194309} (\bibinfo {year} {2014})}\BibitemShut
  {NoStop}%
\bibitem [{\citenamefont {Saremi}(2007)}]{Saremi2007}%
  \BibitemOpen
  \bibfield  {author} {\bibinfo {author} {\bibfnamefont {S.}~\bibnamefont
  {Saremi}},\ }\href {\doibase 10.1103/PhysRevB.76.184430} {\bibfield
  {journal} {\bibinfo  {journal} {Phys. Rev. B}\ }\textbf {\bibinfo {volume}
  {76}},\ \bibinfo {pages} {184430} (\bibinfo {year} {2007})}\BibitemShut
  {NoStop}%
\bibitem [{\citenamefont {Uchoa}\ \emph {et~al.}(2011)\citenamefont {Uchoa},
  \citenamefont {Rappoport},\ and\ \citenamefont {Castro~Neto}}]{Uchoa2011}%
  \BibitemOpen
  \bibfield  {author} {\bibinfo {author} {\bibfnamefont {B.}~\bibnamefont
  {Uchoa}}, \bibinfo {author} {\bibfnamefont {T.~G.}\ \bibnamefont
  {Rappoport}}, \ and\ \bibinfo {author} {\bibfnamefont {A.~H.}\ \bibnamefont
  {Castro~Neto}},\ }\href {\doibase 10.1103/PhysRevLett.106.016801} {\bibfield
  {journal} {\bibinfo  {journal} {Phys. Rev. Lett.}\ }\textbf {\bibinfo
  {volume} {106}},\ \bibinfo {pages} {016801} (\bibinfo {year}
  {2011})}\BibitemShut {NoStop}%
\bibitem [{\citenamefont {Sherafati}\ and\ \citenamefont
  {Satpathy}(2011)}]{Sherafati2011}%
  \BibitemOpen
  \bibfield  {author} {\bibinfo {author} {\bibfnamefont {M.}~\bibnamefont
  {Sherafati}}\ and\ \bibinfo {author} {\bibfnamefont {S.}~\bibnamefont
  {Satpathy}},\ }\href {\doibase 10.1103/PhysRevB.83.165425} {\bibfield
  {journal} {\bibinfo  {journal} {Phys. Rev. B}\ }\textbf {\bibinfo {volume}
  {83}},\ \bibinfo {pages} {165425} (\bibinfo {year} {2011})}\BibitemShut
  {NoStop}%
\bibitem [{\citenamefont {Kirwan}\ \emph {et~al.}(2008)\citenamefont {Kirwan},
  \citenamefont {Rocha}, \citenamefont {Costa},\ and\ \citenamefont
  {Ferreira}}]{Kirwan2008}%
  \BibitemOpen
  \bibfield  {author} {\bibinfo {author} {\bibfnamefont {D.~F.}\ \bibnamefont
  {Kirwan}}, \bibinfo {author} {\bibfnamefont {C.~G.}\ \bibnamefont {Rocha}},
  \bibinfo {author} {\bibfnamefont {A.~T.}\ \bibnamefont {Costa}}, \ and\
  \bibinfo {author} {\bibfnamefont {M.~S.}\ \bibnamefont {Ferreira}},\ }\href
  {\doibase 10.1103/PhysRevB.77.085432} {\bibfield  {journal} {\bibinfo
  {journal} {Phys. Rev. B}\ }\textbf {\bibinfo {volume} {77}},\ \bibinfo
  {pages} {085432} (\bibinfo {year} {2008})}\BibitemShut {NoStop}%
\bibitem [{\citenamefont {Gorman}\ \emph {et~al.}(2015)\citenamefont {Gorman},
  \citenamefont {Duffy}, \citenamefont {Power},\ and\ \citenamefont
  {Ferreira}}]{Gorman2015}%
  \BibitemOpen
  \bibfield  {author} {\bibinfo {author} {\bibfnamefont {P.~D.}\ \bibnamefont
  {Gorman}}, \bibinfo {author} {\bibfnamefont {J.~M.}\ \bibnamefont {Duffy}},
  \bibinfo {author} {\bibfnamefont {S.~R.}\ \bibnamefont {Power}}, \ and\
  \bibinfo {author} {\bibfnamefont {M.~S.}\ \bibnamefont {Ferreira}},\ }\href
  {\doibase 10.1103/PhysRevB.92.035411} {\bibfield  {journal} {\bibinfo
  {journal} {Phys. Rev. B}\ }\textbf {\bibinfo {volume} {92}},\ \bibinfo
  {pages} {035411} (\bibinfo {year} {2015})}\BibitemShut {NoStop}%
\bibitem [{\citenamefont {Xiao}\ \emph {et~al.}(2014)\citenamefont {Xiao},
  \citenamefont {Liu},\ and\ \citenamefont {Wen}}]{Xiao2014}%
  \BibitemOpen
  \bibfield  {author} {\bibinfo {author} {\bibfnamefont {X.}~\bibnamefont
  {Xiao}}, \bibinfo {author} {\bibfnamefont {Y.}~\bibnamefont {Liu}}, \ and\
  \bibinfo {author} {\bibfnamefont {W.}~\bibnamefont {Wen}},\ }\href {\doibase
  10.1088/0953-8984/26/26/266001} {\bibfield  {journal} {\bibinfo  {journal}
  {J. Phys.: Condens. Matter}\ }\textbf {\bibinfo {volume} {26}},\ \bibinfo
  {pages} {266001} (\bibinfo {year} {2014})}\BibitemShut {NoStop}%
\bibitem [{\citenamefont {Zare}\ \emph {et~al.}(2016)\citenamefont {Zare},
  \citenamefont {Parhizgar},\ and\ \citenamefont {Asgari}}]{Zare2016}%
  \BibitemOpen
  \bibfield  {author} {\bibinfo {author} {\bibfnamefont {M.}~\bibnamefont
  {Zare}}, \bibinfo {author} {\bibfnamefont {F.}~\bibnamefont {Parhizgar}}, \
  and\ \bibinfo {author} {\bibfnamefont {R.}~\bibnamefont {Asgari}},\ }\href
  {\doibase 10.1103/PhysRevB.94.045443} {\bibfield  {journal} {\bibinfo
  {journal} {Phys. Rev. B}\ }\textbf {\bibinfo {volume} {94}},\ \bibinfo
  {pages} {045443} (\bibinfo {year} {2016})}\BibitemShut {NoStop}%
\bibitem [{\citenamefont {Patrone}\ and\ \citenamefont
  {Einstein}(2012)}]{Patrone2012}%
  \BibitemOpen
  \bibfield  {author} {\bibinfo {author} {\bibfnamefont {P.~N.}\ \bibnamefont
  {Patrone}}\ and\ \bibinfo {author} {\bibfnamefont {T.~L.}\ \bibnamefont
  {Einstein}},\ }\href {\doibase 10.1103/PhysRevB.85.045429} {\bibfield
  {journal} {\bibinfo  {journal} {Phys. Rev. B}\ }\textbf {\bibinfo {volume}
  {85}},\ \bibinfo {pages} {045429} (\bibinfo {year} {2012})}\BibitemShut
  {NoStop}%
\bibitem [{\citenamefont {Bollinger}\ \emph {et~al.}(2001)\citenamefont
  {Bollinger}, \citenamefont {Lauritsen}, \citenamefont {Jacobsen},
  \citenamefont {N\o{}rskov}, \citenamefont {Helveg},\ and\ \citenamefont
  {Besenbacher}}]{Bollinger2001}%
  \BibitemOpen
  \bibfield  {author} {\bibinfo {author} {\bibfnamefont {M.~V.}\ \bibnamefont
  {Bollinger}}, \bibinfo {author} {\bibfnamefont {J.~V.}\ \bibnamefont
  {Lauritsen}}, \bibinfo {author} {\bibfnamefont {K.~W.}\ \bibnamefont
  {Jacobsen}}, \bibinfo {author} {\bibfnamefont {J.~K.}\ \bibnamefont
  {N\o{}rskov}}, \bibinfo {author} {\bibfnamefont {S.}~\bibnamefont {Helveg}},
  \ and\ \bibinfo {author} {\bibfnamefont {F.}~\bibnamefont {Besenbacher}},\
  }\href {\doibase 10.1103/PhysRevLett.87.196803} {\bibfield  {journal}
  {\bibinfo  {journal} {Phys. Rev. Lett.}\ }\textbf {\bibinfo {volume} {87}},\
  \bibinfo {pages} {196803} (\bibinfo {year} {2001})}\BibitemShut {NoStop}%
\bibitem [{\citenamefont {Pavlovi\ifmmode~\acute{c}\else \'{c}\fi{}}\ and\
  \citenamefont {Peeters}(2015)}]{Pavlovic2015}%
  \BibitemOpen
  \bibfield  {author} {\bibinfo {author} {\bibfnamefont {S.}~\bibnamefont
  {Pavlovi\ifmmode~\acute{c}\else \'{c}\fi{}}}\ and\ \bibinfo {author}
  {\bibfnamefont {F.~M.}\ \bibnamefont {Peeters}},\ }\href {\doibase
  10.1103/PhysRevB.91.155410} {\bibfield  {journal} {\bibinfo  {journal} {Phys.
  Rev. B}\ }\textbf {\bibinfo {volume} {91}},\ \bibinfo {pages} {155410}
  (\bibinfo {year} {2015})}\BibitemShut {NoStop}%
\bibitem [{\citenamefont {Segarra}\ \emph {et~al.}(2016)\citenamefont
  {Segarra}, \citenamefont {Planelles},\ and\ \citenamefont
  {Ulloa}}]{Segarra2016}%
  \BibitemOpen
  \bibfield  {author} {\bibinfo {author} {\bibfnamefont {C.}~\bibnamefont
  {Segarra}}, \bibinfo {author} {\bibfnamefont {J.}~\bibnamefont {Planelles}},
  \ and\ \bibinfo {author} {\bibfnamefont {S.~E.}\ \bibnamefont {Ulloa}},\
  }\href {\doibase 10.1103/PhysRevB.93.085312} {\bibfield  {journal} {\bibinfo
  {journal} {Phys. Rev. B}\ }\textbf {\bibinfo {volume} {93}},\ \bibinfo
  {pages} {085312} (\bibinfo {year} {2016})}\BibitemShut {NoStop}%
\bibitem [{\citenamefont {Farmanbar}\ \emph {et~al.}(2016)\citenamefont
  {Farmanbar}, \citenamefont {Amlaki},\ and\ \citenamefont
  {Brocks}}]{Farmanbar2016}%
  \BibitemOpen
  \bibfield  {author} {\bibinfo {author} {\bibfnamefont {M.}~\bibnamefont
  {Farmanbar}}, \bibinfo {author} {\bibfnamefont {T.}~\bibnamefont {Amlaki}}, \
  and\ \bibinfo {author} {\bibfnamefont {G.}~\bibnamefont {Brocks}},\ }\href
  {\doibase 10.1103/PhysRevB.93.205444} {\bibfield  {journal} {\bibinfo
  {journal} {Phys. Rev. B}\ }\textbf {\bibinfo {volume} {93}},\ \bibinfo
  {pages} {205444} (\bibinfo {year} {2016})}\BibitemShut {NoStop}%
\bibitem [{\citenamefont {Rostami}\ \emph {et~al.}(2016)\citenamefont
  {Rostami}, \citenamefont {Asgari},\ and\ \citenamefont
  {Guinea}}]{Rostami2016}%
  \BibitemOpen
  \bibfield  {author} {\bibinfo {author} {\bibfnamefont {H.}~\bibnamefont
  {Rostami}}, \bibinfo {author} {\bibfnamefont {R.}~\bibnamefont {Asgari}}, \
  and\ \bibinfo {author} {\bibfnamefont {F.}~\bibnamefont {Guinea}},\ }\href
  {\doibase 10.1088/0953-8984/28/49/495001} {\bibfield  {journal} {\bibinfo
  {journal} {J. Phys.: Condens. Matter}\ }\textbf {\bibinfo {volume} {28}},\
  \bibinfo {pages} {495001} (\bibinfo {year} {2016})}\BibitemShut {NoStop}%
\bibitem [{\citenamefont {Liu}\ \emph {et~al.}(2013)\citenamefont {Liu},
  \citenamefont {Shan}, \citenamefont {Yao}, \citenamefont {Yao},\ and\
  \citenamefont {Xiao}}]{Liu2013}%
  \BibitemOpen
  \bibfield  {author} {\bibinfo {author} {\bibfnamefont {G.-B.}\ \bibnamefont
  {Liu}}, \bibinfo {author} {\bibfnamefont {W.-Y.}\ \bibnamefont {Shan}},
  \bibinfo {author} {\bibfnamefont {Y.}~\bibnamefont {Yao}}, \bibinfo {author}
  {\bibfnamefont {W.}~\bibnamefont {Yao}}, \ and\ \bibinfo {author}
  {\bibfnamefont {D.}~\bibnamefont {Xiao}},\ }\href {\doibase
  10.1103/PhysRevB.88.085433} {\bibfield  {journal} {\bibinfo  {journal} {Phys.
  Rev. B}\ }\textbf {\bibinfo {volume} {88}},\ \bibinfo {pages} {085433}
  (\bibinfo {year} {2013})}\BibitemShut {NoStop}%
\bibitem [{\citenamefont {Deaven}\ \emph {et~al.}(1991)\citenamefont {Deaven},
  \citenamefont {Rokhsar},\ and\ \citenamefont {Johnson}}]{Deaven1991}%
  \BibitemOpen
  \bibfield  {author} {\bibinfo {author} {\bibfnamefont {D.~M.}\ \bibnamefont
  {Deaven}}, \bibinfo {author} {\bibfnamefont {D.~S.}\ \bibnamefont {Rokhsar}},
  \ and\ \bibinfo {author} {\bibfnamefont {M.}~\bibnamefont {Johnson}},\ }\href
  {\doibase 10.1103/PhysRevB.44.5977} {\bibfield  {journal} {\bibinfo
  {journal} {Phys. Rev. B}\ }\textbf {\bibinfo {volume} {44}},\ \bibinfo
  {pages} {5977} (\bibinfo {year} {1991})}\BibitemShut {NoStop}%
\bibitem [{Note1()}]{Note1}%
  \BibitemOpen
  \bibinfo {note} {Notice that we use capital letters for the spin direction in
  order to avoid confusion with the notation for orbitals.}\BibitemShut {Stop}%
\bibitem [{\citenamefont {Mattis}(2006)}]{Mattis}%
  \BibitemOpen
  \bibfield  {author} {\bibinfo {author} {\bibfnamefont {D.~C.}\ \bibnamefont
  {Mattis}},\ }\href@noop {} {\emph {\bibinfo {title} {The theory of magnetism
  made simple}}}\ (\bibinfo  {publisher} {World Scientific},\ \bibinfo
  {address} {Singapore},\ \bibinfo {year} {2006})\BibitemShut {NoStop}%
\bibitem [{\citenamefont {Nolting}\ and\ \citenamefont
  {Ramakanth}(2009)}]{Nolting}%
  \BibitemOpen
  \bibfield  {author} {\bibinfo {author} {\bibfnamefont {W.}~\bibnamefont
  {Nolting}}\ and\ \bibinfo {author} {\bibfnamefont {A.}~\bibnamefont
  {Ramakanth}},\ }\href@noop {} {\emph {\bibinfo {title} {Quantum Theory of
  Magnetism}}}\ (\bibinfo  {publisher} {Springer},\ \bibinfo {address} {Berlin
  Heidelberg},\ \bibinfo {year} {2009})\BibitemShut {NoStop}%
\end{thebibliography}%

\end{document}